\documentclass[acmsmall]{acmart} 

\usepackage{csquotes}
\usepackage{hyperref,cleveref}
\usepackage{mdframed}
\usepackage{multirow}
\usepackage{xspace}

\setcopyright{acmlicensed}
\copyrightyear{2025}
\acmYear{2025}
\acmDOI{10.1145/3728638}

\acmJournal{CSUR}
\acmVolume{}
\acmNumber{}
\acmArticle{}
\acmMonth{4}

\newcommand{\eg}{e.g.,\xspace}
\newcommand{\ie}{i.e.,\xspace}

\newcommand{\etal}{\textit{et al}.\xspace}

\definecolor{headergray}{gray}{0.7}
\definecolor{rowgray}{gray}{0.88}

\definecolor{guidelines}{HTML}{B8E1FF}
\definecolor{refmodels}{HTML}{86B9E9}
\definecolor{measurement}{HTML}{85CAEA}
\definecolor{techniques}{HTML}{51B4E1}

\newenvironment{highlights}{
  \begin{center}
    \begin{tabular}{|p{.02\textwidth}p{.92\textwidth}|}
      \hline
        \multicolumn{2}{|l|}{\textbf{Highlights}} \\
        \hline
}{
      \hline
    \end{tabular}
  \end{center}
}
\newcommand{\highlight}[2]{$H_{#1}$ & #2. \\}

\newenvironment{revenv}{}{} 
\newcommand{\rev}[1]{#1} 

\begin{document}

\title{\rev{Carbon-Efficient} Software Design and Development: A Systematic Literature Review}

\author{Ornela Danushi}\authornote{Author names are alphabetically sorted. All Authors contributed equally to this study. This is the authors' version of the work. It is posted here for your personal use. Not for redistribution. The definitive Version of Record was published in {\it ACM Computing Surveys}, \url{https://doi.org/10.1145/3728638}.}
\affiliation{%
  \institution{Department of Computer Science, University of Pisa}
  \city{Pisa}
  \country{Italy}
}
\email{ornela.danushi@phd.unipi.it}
\author{Stefano Forti}\authornotemark[1]
\affiliation{%
  \institution{Department of Computer Science, University of Pisa}
  \city{Pisa}
  \country{Italy}
}
\email{stefano.forti@unipi.it}
\author{Jacopo Soldani}\authornotemark[1]
\affiliation{%
  \institution{Department of Computer Science, University of Pisa}
  \city{Pisa}
  \country{Italy}
}
\email{jacopo.soldani@unipi.it}

\renewcommand{\shortauthors}{Danushi et al.}

\begin{abstract}
The ICT sector, responsible for 2\% of \rev{global carbon emissions}, 
is under scrutiny calling for methodologies and tools to design and develop software in an environmentally sustainable-by-design manner. 
However, the software engineering solutions for designing and developing \rev{carbon-efficient} software are currently scattered over multiple different pieces of literature, which makes it difficult to consult the body of knowledge on the topic.
In this article, we precisely conduct a systematic literature review on state-of-the-art proposals for designing and developing \rev{carbon-efficient} software. We identify and analyse 65 primary studies by classifying them through a taxonomy aimed at answering the 5W1H questions of \rev{carbon-efficient} software design and development. We first provide a reasoned overview and discussion of the existing guidelines, reference models, measurement solutions and techniques for measuring, reducing, or minimising the 
\rev{carbon footprint} of software. Ultimately, we identify open challenges and research gaps, offering insights for future work in this field.
\end{abstract}



\received{29 July 2024}
\received[revised]{24 February 2025}
\received[accepted]{01 April 2025}
\maketitle

\section{Introduction}
\label{sec:introduction}
In recent years, the themes of \textit{sustainability} and \textit{sustainable development} have seen a growing interest from governments, academia and industry, aiming at balancing the need to safeguard our planet along with social and economical instances~\cite{brundtland1987our,sachs2022sustainable}. Such effort is clearly epitomised by the adoption of the \textit{Sustainable Development Goals} in the UN action plan reported in the \textit{2030 Agenda for Sustainable Development}.\footnote{Publicly available at: \url{https://sdgs.un.org/2030agenda}} Indeed, global carbon emissions need to be reduced to secure our Planet’s future by achieving the ambition set by the Paris agreement\footnote{https://www.un.org/en/climatechange/paris-agreement} \rev{\cite{Manner2022_BlackSoftware,Moeller2024_ZeroGHG}}. However, such an objective remains far from being reached, as highlighted by the outcomes of the \textit{2023 United Nations Climate Change Conference} (COP28).

Building on this momentum, there is a rising concern about measuring, assessing, and improving the sustainability of Information and Communication Technology (ICT). Indeed, ICT is currently responsible for 2\% of global carbon emissions \rev{\cite{Manner2022_BlackSoftware,Mingay2008_ICTEmissions}}, with an energy consumption estimated at between 6\% and 9\% of the global demand and expected to grow up to 20\% by 2030~\cite{belkhirAssessing2018,carbonict2023,ADISA2024141768}.
Most efforts to reduce the carbon footprint of ICT were dedicated to improve hardware energy-efficiency, by leveraging the combined effects of Moore's law~\cite{Moore98a} and Dennard scaling~\cite{dennard1974}. Moore's law ensured the possibility of packing double the transistors in a \rev{chip}, while Dennard scaling has ensured that integrated circuits of the same area maintain constant power consumption regardless of the increased number of transistors. This resulted in faster chips with no increase in energy consumption up until the 2010s~\cite{anderson2023treehouse}. 

With Dennard's scaling and Moore's law reaching their limits~\cite{DBLP:journals/csur/MuralidharBB22}, improving the environmental sustainability of ICT will necessarily involve making software more sustainable by 
\rev{reducing the \textit{carbon footprint}} associated with its deployment. On this line, much research has targeted the optimisation of renewable energy usage through suitable service orchestration and consolidation~\cite{GaglianeseGreenOrch2023}, without considering the potential of (re)designing and (re)writing software according to a sustainable-by-design approach. Naturally, as pointed out by Anderson et al.~\cite{anderson2023treehouse}, further improvements in software sustainability will require treating 
\rev{carbon-efficiency} in our code design and development. Similarly, Verdecchia et al.~\cite{DBLP:journals/software/VerdecchiaLEVE21} call for methodologies to write sustainable software by blending decision-making with the software execution context.
Although the term green software engineering was coined more than a decade ago~\cite{weneedtodecarbonise}, much remains to be done to fully rethink how \rev{carbon-efficient} software is designed and implemented. 

\begin{revenv}
At the same time, as observed, \eg by the Green Software Foundation \cite{GSF2021_10Recs} and Kern \etal \cite{61_Kern2015_ICTCO2footprint}, the carbon-efficiency of software products is mostly affected by their \textit{design}, \textit{development}, and \textit{operation}.
Indeed, while sustainability considerations span the entire software lifecycle, design, development, and operation represent the primary leverage points for reducing software carbon emissions, \eg through software architecture decisions, efficient data structures and algorithms, and energy- and carbon-aware orchestration strategies.
\end{revenv}

\rev{While the environmentally sustainable operation of software products has already been analysed, \eg by Gaglianese \etal  \cite{Gaglianese2023_GreenCloudEdgeOrchestration}, this} article precisely aims at surveying the state of the art related to the design and development of \rev{carbon-efficient} software. 
Currently available \rev{methodologies} are scattered over multiple different pieces of literature. Unfortunately, this makes it difficult to consult the body of knowledge on the topic, both for researchers willing to investigate \rev{carbon-efficient software design and development} methodologies, and for practitioners willing to apply them to their software systems. To achieve this objective, this article conducts a systematic literature review (SLR), which identifies, taxonomically classifies, and compares the existing guidelines, reference models, measurement solutions, and techniques for designing and developing \rev{carbon-efficient} software.
Following the guidelines by Petersen et al.~\cite{Petersen2008_SMSGuidelines,Petersen2015_SMSGuidelinesUpdate} on how to conduct SLRs in software engineering, we first identify a corpus of 65 primary \rev{studies}. We then classify and compare the selected studies by means of a newly proposed taxonomy, aimed at highlighting the \textit{5W1H} of \rev{carbon-efficient software design and development}, viz., \textit{who} are the stakeholders targeted by the selected studies, \textit{what} methods are proposed therein, \textit{why}, \textit{where} and \textit{when} the proposed methods are applied (in terms of application domain and software lifecycle step, respectively), and \textit{how} the selected studies have been released to the community.

Overall, our SLR pursues a twofold ultimate objective. On one hand, it aims at accurately depicting existing proposals, methodologies, and solutions that enable the design and development of environmentally sustainable software, namely for measuring, reducing, and/or minimising a software’s \rev{carbon emissions}. On the other hand, it analyses the state of the art in \rev{carbon-efficient software design and development} with the aim of shedding light on open challenges and gaps, which can contribute to reducing the carbon footprint of ICT from a software perspective.

\medskip\noindent The rest of this article is organised as follows.  
\Cref{sec:design} describes the design of the SLR process.
\Cref{sec:sota} classifies and summarises the state of the art on \rev{carbon-efficient} software design and development.
The classification is retaken in \Cref{sec:5w1h}, which provides an analysis of the 5W1H of designing and developing \rev{carbon-efficient} software.
\Cref{sec:ttv,sec:related} discuss the possible threats to the validity of our SLR and related work, respectively.
Finally, \Cref{sec:conclusions} concludes the article by highlighting future research directions for the design and development of \rev{carbon-efficient} software.

\section{Research Design}
\label{sec:design}
\begin{figure}
    \centering
    \includegraphics[width=.95\textwidth]{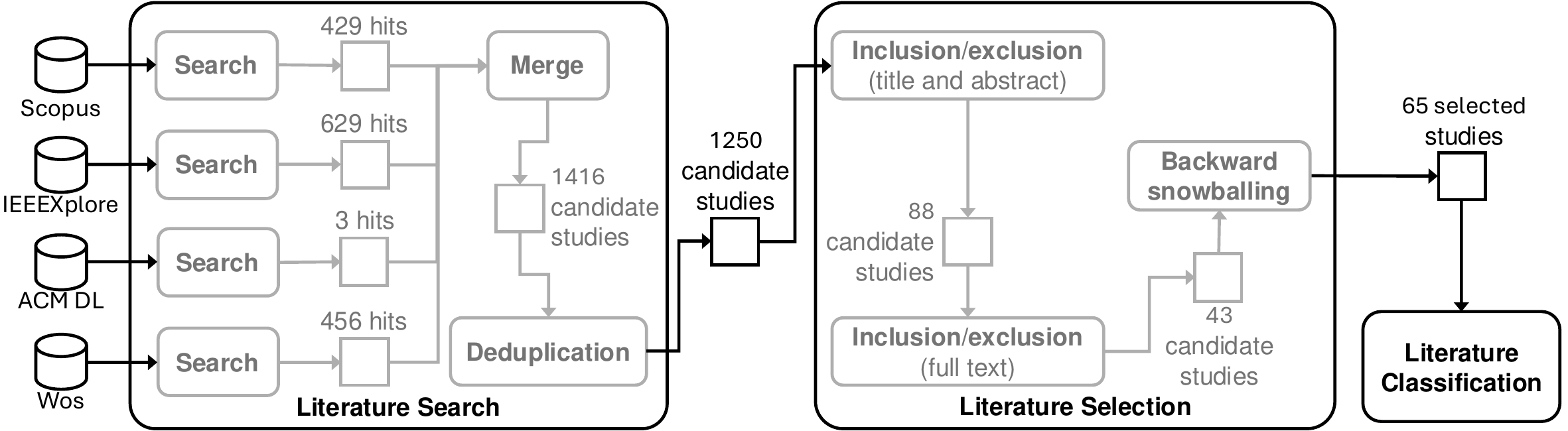} 
    \caption{SLR process for identifying and classifying existing literature of interest.} 
    \label{fig:design:workflow}
\end{figure}

After explicitly stating the objective and the 5W1H questions that drove our SLR (\Cref{sec:design:rq}), we illustrate the design of the SLR (\Cref{sec:design:search,sec:design:selection,sec:design:classification}) by relying on \Cref{fig:design:workflow} to provide a graphical sketch of the SLR process. 
We also provide a replication package for repeating our SLR (\Cref{sec:design:replication-package}).

\subsection{The 5W1H Questions}
\label{sec:design:rq}

The objective of this article is to identify, classify, and analyse the state of the art in the design and development of environmentally sustainable software by limiting its \rev{carbon footprint}.
This is done by answering to the 5W1H questions listed below:

\smallskip \noindent 
\textbf{Who?} Who are the stakeholders in existing techniques for \rev{carbon-efficient} software design and development?

\smallskip \noindent 
\textbf{What?} What methods, tools, and techniques have been proposed to achieve \rev{carbon-efficient} software design and development? 

\smallskip \noindent 
\textbf{Why?} Why does a selected study target the \rev{carbon efficiency} of software? Does it aim at \rev{measuring, reducing, or optimising its carbon footprint}?

\smallskip \noindent 
\textbf{Where?} In which application domains can a selected study be applied?

\smallskip \noindent 
\textbf{When?} In which software lifecycle stages can a selected study's proposal be used?
    
\smallskip \noindent 
\textbf{How?} How were the selected studies published?

\subsection{Literature Search}
\label{sec:design:search}

Following the guidelines by Petersen et al. \cite{Petersen2008_SMSGuidelines,Petersen2015_SMSGuidelinesUpdate}, we identified the search string guided by our main research objective, namely to overview the state of the art in \rev{carbon-efficient} software design and development.
More precisely, we defined the search strings based on the PICO (\textit{Population, Intervention, Comparison, Outcome}) terms characterising our research questions \cite{Kitchenham2007_GuidelinesSLRinSE}, taking keywords from each aspect of our research objective -- and the corresponding 5W1H questions. 
Differently from Petersen et al. \cite{Petersen2008_SMSGuidelines,Petersen2015_SMSGuidelinesUpdate}, we did not limit our focus to specific outcomes or experimental designs in our study.
Rather, we aimed at providing a broader overview of the state of the art in the design and development of \rev{carbon-efficient} software. 
This difference is also reflected in the search string we employed to search for studies, \ie
\begin{footnotesize}
    \begin{align}
        & (\texttt{carbon} \vee \texttt{co2} \vee \texttt{emission*}) \wedge \\
        & (\texttt{aware*} \vee \texttt{contain*} \vee \texttt{reduc*} \vee \texttt{optim*} \vee \texttt{minim*} \vee \texttt{assess*}) \wedge \\
        & (\texttt{software design*} \vee \texttt{design pattern*} \vee \texttt{software develop* } \vee &\notag\\
        &   \texttt{software engineer*} \vee \texttt{software cod*} \vee \texttt{software program*})
    \end{align}
\end{footnotesize}
(where “{\tt\small *}” matches lexically related terms, e.g., plurals and conjugations, and relevant synonyms).
The search string is essentially the conjunction of three groups, aimed at identifying the existing efforts (1)~considering carbon emissions, (2)~reducing or containing such carbon emissions, and (3)~focusing on software design and development.

We used the aforementioned string to search for \rev{all white literature published before January 2024, when this research began}.
The search was carried on ACM Digital Library, IEEEXplore, Scopus, and Web of Science, \rev{and was restricted to white literature published in Computer Science venues}.
We retrieved a total of \rev{1517} candidate studies, which -- after checking for duplicates -- were reduced to a list of \rev{1273} unique candidate studies.

\subsection{Literature Selection}
\label{sec:design:selection}
The search terms used in \Cref{sec:design:search} were checked against the titles and abstracts of the existing literature, which resulted in a significant number of candidate studies that were not relevant for our purposes.
Therefore, the list of candidate studies was further refined by relying on the inclusion/exclusion criteria in \Cref{tab:design:ie-criteria}.
The inclusion criteria $i_1$ and $i_2$ aligned the selection with our main research objective by allowing the selection of existing studies that support the design and/or development of software, while also aiming at reducing its carbon footprint.
The exclusion criteria $e_1$, $e_2$, and $e_3$ restricted the focus to peer-reviewed literature written in English, which was also accessible from our institution.
Finally, the exclusion criteria $e_4$ and $e_5$ restricted the focus to primary studies and excluding collections of papers and secondary studies, respectively.
A candidate study was selected if satisfying \textit{all} the inclusion criteria, whereas it was excluded if satisfying \textit{any} of the exclusion criteria.
\begin{table}[t]
    \scriptsize
    \caption{Inclusion and exclusion criteria.}
    \label{tab:design:ie-criteria}
    \begin{tabular}{ccp{.78\textwidth}}
        \hline 
        \rowcolor{headergray}
        \textbf{Criteria} 
            & \textbf{Type}
            & \textbf{Description}
            \\
        \hline
        $i_1$    
            & Inclusion
            & The study explicitly targets reducing the carbon footprint of software
            \\ 
        \rowcolor{rowgray}
        $i_2$   
            & Inclusion    
            & The study proposes solutions to support the design and/or development of software
            \\
        $e_1$
            & Exclusion    
            & The study is not peer-reviewed  
            \\
        \rowcolor{rowgray}
        $e_2$
            & Exclusion    
            & The study is not written in English 
            \\
        $e_3$
            & Exclusion    
            & The study is not accessible from our institution 
            \\ 
        \rowcolor{rowgray}
        $e_4$
            & Exclusion    
            & The study is \rev{an edited volume collecting multiple research papers (\eg conference proceedings or edited books) rather than a single, cohesive research work.}
            \\ 
        $e_5$
            & Exclusion    
            & The study is a secondary study (\eg a survey or SLR) 
            \\ 
        \hline 
    \end{tabular}
\end{table}

The inclusion/exclusion criteria were applied in two consecutive steps.
%
In both steps, each candidate study was marked as to be selected or excluded by two authors independently.
If there was disagreement about the inclusion of a candidate study, the author who was not involved in the coding of the study acted as a third evaluator to resolve the disagreement.
In the first step, the selection was based on assessing the inclusion/exclusion criteria over the title and abstract of candidate studies, which reduced the list of candidate studies to 88 candidate studies, with an overall inter-rater agreement of \rev{85.6}\%.\footnote{We measured the inter-rater agreement on the inclusion/exclusion coding by exploiting the Krippendorff-$\alpha$ coefficient~\cite{Krippendorff2004_ContentAnalysis}, which classifies a coding as good if the value is greater or equal to 80\%.}
The process was then repeated by considering the full text of the remaining 88 candidate studies, which further narrowed the list to 43 candidate studies, with an overall inter-rater agreement of 90.9\%.

\begin{table}
    \caption{List of selected studies, sorted by reference number.}
    \label{tab:design:selected-studies}
    \scriptsize
\begin{tabular}{p{.99\textwidth}}
    \hline
    \rowcolor{headergray}
    \textbf{Ref.} \textbf{Author(s). \textit{Title}} (\textbf{Year})
        \\
        \hline
        \cite{16_Ahmed2020_codingCloud} Ahmed. \textit{Environmental sustainability coding techniques for cloud computing} (2020) \\ 
        \rowcolor{rowgray}
        \cite{31_Alofi2023_Self-Optimizing} Alofi et al. \textit{Self-optimizing the environmental sustainability of blockchain-based systems} (2023) \\ 
        \cite{30_Alofi2022_OptimizingConsensus} Alofi et al. \textit{Optimizing the energy consumption of blockchain-based systems using evolutionary algorithms: a new problem ...} 
        (2022) \\ 
        \rowcolor{rowgray}
        \cite{18_Atkinson2014_GreenSpecifications} Atkinson et al. \textit{Facilitating greener IT through green specifications} (2014) \\ 
        \cite{50_Becker2016_Requirements} Becker et al. \textit{Requirements: the key to sustainability} (2016) \\ 
        \rowcolor{rowgray}
        \cite{21_Beghoura2014_GreenEvaluation} Beghoura et al. \textit{Green application awareness: nonlinear energy consumption model for green evaluation} (2014) \\ 
        \cite{17_Brownlee2021_AccuracyEnergyML} Brownleee et al. \textit{Exploring the accuracy-energy trade-off in machine learning} (2021) \\ 
        \rowcolor{rowgray}
        \cite{01_Chauhan2013_GreenSDLCCloud} Chauan and Saxena. \textit{A green software development life cycle for cloud computing} (2013) \\
        \cite{36_Fernandez2018_GreenDesign} Condori-Fernàndez and Lago. \textit{The influence of green strategies design onyo quality requirement prioritization} (2018) \\ 
        \rowcolor{rowgray}
        \cite{24_Dick2013_GSE-Agile} Dick et al. \textit{Green software engineering with agile methods} (2013) \\ 
        \cite{48_Dick2010_GreenSustainableSoftware} Dick et al. \textit{A model and selected instances of green and sustainable software} (2010) \\ 
        \rowcolor{rowgray}
        \cite{32_Fakhar2012_GreenComputing} Fakhar et al. \textit{Software level green computing for large scale systems} (2012) \\ 
        \cite{52_Fu2021_ComputerVision} Fu et al. \textit{Reconsidering CO2 emissions from computer vision} (2021) \\ 
        \rowcolor{rowgray}
        \cite{51_LeGoaer2021_greenCode} Le Goaër. \textit{Enforcing green code with Android lint} (2021) \\ 
        \cite{13_LeGoaer2023_ecoCodeProject} Le Goaër. \textit{Decarbonizinf software with free and open-source software: the ecoCode project} (2023) \\ 
        \rowcolor{rowgray}
        \cite{49_LeGoaer2023_ecoCode} Le Goaër. \textit{ecoCode: a SonarQube plugin to remove energy smells from Android projects} (2023) \\ 
        \cite{15_Heithoff2023_DigitalTwins} Heithoff et al. \textit{Digital twins for sustainable software systems} (2023) \\ 
        \rowcolor{rowgray}
        \cite{39_Hoesch-Klohe2010_GreenBPM} Hoesch-Klohe et al. \textit{Towards green business process management} (2010) \\ 
        \cite{34_Ibrahim2021_GSDProcessModel} Ibrahim et al. \textit{The development of green software process model: a qualitative design and pilot study} (2021) \\ 
        \rowcolor{rowgray}
        \cite{25_Ibrahim2022_GreenFactors} Ibrahim et al. \textit{Green software process factors: a qualitative study} (2022) \\ 
        \cite{07_Jayanthi20211_OrganizationalStructure} Jayanthi et al. \textit{An organizational structure for sustainable software development} (2021) \\ 
        \rowcolor{rowgray}
        \cite{26_Jebraeil2017_gUML} Jebraeil et al. \textit{gUML: Reasoning about energy at design time by extending UML deployment diagrams with data centre contextual ...} 
        (2017) \\ 
        \cite{46_Johann2011_IntegratedApproach} Johann et al. \textit{Sustainable development, sustainable software, and sustainable software engineering: an integrated approach} (2011) \\ 
        \rowcolor{rowgray}
        \cite{03_Karunakaran2013_PetriNet} Karunakaran and Rao. \textit{A petri net simulation of software development lifecycle towards green IT} (2013) \\ 
        \cite{54_Kern2013_GreenSA} Kern et al. \textit{Green software and green software engineering -- definitions, measurements, and quality aspects} (2013) \\ 
        \rowcolor{rowgray}
        \cite{61_Kern2015_ICTCO2footprint} Kern et al. \textit{Impacts of software and its engineering on the carbon footprint of ICT} (2015) \\ 
        \cite{33_Kern2018_AssessCriteria} Kern et al. \textit{Sustainable software products -- Towards assessment criteria for resource and energy efficiency} (2018) \\ 
        \rowcolor{rowgray}
        \cite{64_Kharchenko2013_VonNeumannParadigm} Kharchenko and Gorbenko. \textit{Evolution of von Neumann's paradigm: dependable and green computing} (2013) \\ 
        \cite{62_Kipp2011_GreenMetrics} Kipp et al. \textit{Green metrics for energy-aware IT systems} (2011) \\ 
        \rowcolor{rowgray}
        \cite{28_Kipp2012_GreenIndicators} Kipp et al. \textit{Layered green performance indicators} (2012) \\ 
        \cite{65_Kocak2019_UtilityModel} Kocak and Alptekin. \textit{A utility model for designing environmentally sustainable software} (2019) \\ 
        \rowcolor{rowgray}
        \cite{57_Lago2019_DecisionMaps} Lago. \textit{Architecture design decision maps for software sustainability} (2019) \\ 
        \cite{11_Lago2011_SOSE} Lago and Jansen. \textit{Creating environmental awareness in service oriented software engineering} (2011) \\ 
        \rowcolor{rowgray}
        \cite{44_Lami2012_SoftwareMeasurement} Lami and Buglione. \textit{Measuring software sustainability from a process-centric perspective} (2012) \\ 
        \cite{06_Lami2014_MeasurementFramework} Lami et al. \textit{An ISO/IEC 33000-compliant measurement framework for software process sustainability assessment} (2014) \\ 
        \rowcolor{rowgray}
        \cite{02_Lami2013_SustainabilityIndicators} Lami et al. \textit{A methodology to derive sustainability indicators for software development projects} (2013) \\
        \cite{59_Lasla2022_GreenPoWAE} Lasla et al. \textit{Green-PoW: an energy-efficient blockchain proof-of-work consensus algorithm} (2022) \\ 
        \rowcolor{rowgray}
        \cite{58_Mahmoud2013_GreenModel} Mahmoud and Ahmad. \textit{A green model for sustainable software engineering} (2013) \\ 
        \cite{19_Mancebo2021_FEETINGS} Mancebo et al. \textit{FEETINGS: Framework for energy efficiency testing to improve environmental goals of the software} (2021) \\ 
        \rowcolor{rowgray}
        \cite{08_Mehra2023_AssessRefactoring} Mehra et al. \textit{Assessing the impact of refactoring energy-inefficient code patterns on software sustainability: an industry case study} (2023) \\
        \cite{38_Sharma2022_GreenQuotient} Mehra et al. \textit{Towards a green quotient for software projects} (2022) \\ 
        \rowcolor{rowgray}
        \cite{56_Mohankumar2016_GSDLCM} Mohankumar and Kumar. \textit{Green based software development life cycle model for software engineering} (2016) \\ 
        \cite{05_Moshnyaga2013_LCA} Moshnyaga. \textit{An assessment of software lifecycle energy} (2013) \\
        \rowcolor{rowgray}
        \cite{09_Moshnyaga2013_AssessLContribution} Moshnyaga. \textit{Assessment of software lifecycle energy and its contribution to green house gas emissions} (2013) \\ 
        \cite{29_Moshnyaga2013_AssessLC} Moshnyaga. \textit{Lifecycle energy assessment of mobile applications} (2013) \\ 
        \rowcolor{rowgray}
        \cite{45_Naumann2011_GREENSOFT} Naumann et al. \textit{The GREENSOFT model: a reference model for green and sustainable software and its engineering} (2011) \\ 
        \cite{60_Naumann2015_QualityModels} Naumann et al. \textit{Sustainable software engineering: process and quality models, life cycle, and social aspects} (2015) \\ 
        \rowcolor{rowgray}
        \cite{04_Noureddine2012_GREENS} Noureddine et al. \textit{A preliminary study of the impact of software engineering on GreenIT} (2012) \\
        \cite{12_Pa2017_GreenDesignMeasurement} Pa et al. \textit{Dashboard system for measuring green software design} (2021) \\ 
        \rowcolor{rowgray}
        \cite{14_Pan2022_CNNDecomposition} Pan and Rajan. \textit{Decomposing convolutional neural networks into reusable and replaceable modules} (2022) \\ 
        \cite{53_Patterson2022_MLCO2footprint} Patterson et al. \textit{The carbon footprint of machine learning will plateau, then shrink} (2022) \\ 
        \rowcolor{rowgray}
        \cite{63_Penzenstadler2013_GenericModel} Penzenstadler and Femmer. \textit{A generic model for sustainability with process-- and product-specific instances} (2013) \\ 
        \cite{10_Ponsard2018_CaseStudy} Ponsard et al. \textit{Building sustainable software for sustainable systems: case study of a shared pick-up and delivery service} (2018) \\
        \rowcolor{rowgray}
        \cite{22_Radersma2022_GreenEvaluation} Radersma. \textit{Green coding: reduce your carbon footprint} (2022) \\ 
        \cite{35_Raisian2021_MeasurementStructure} Raisian et al. \textit{The green software measurement structure based on sustainability perspective} (2021) \\ 
        \rowcolor{rowgray}
        \cite{47_Sharma2015_GreenSDLC} Sharma et al. \textit{Energy efficient software development life cycle -- an approach towards smart computing} (2015) \\ 
        \cite{55_Shenoy2011_GreenDevelopmentModel} Shenoy et al. \textit{Green software development model: an approach towards sustainable software development} (2011) \\ 
        \rowcolor{rowgray}
        \cite{23_Siegmund2022_GreenConfiguration} Siegmund et al. \textit{Green configuration: can artificial intelligence help reduce energy consumption of configurable software systems?} (2022) \\ 
        \cite{20_Sikanda2023_GreenAIQuotient} Sikand et al. \textit{Green AI quotient: assessing greenness of AI-based software and the way forward} (2023) \\ 
        \rowcolor{rowgray}
        \cite{42_Simon2023_ImpactSLC} Simon et al. \textit{Uncovering the environmental impact of software life cycle} (2023) \\ 
        \cite{37_Tee2014_GreenKnowledge} Tee et al. \textit{Toward developing green software development model in managing knowledge of IT practitioners for sustaining future...} 
        (2014) \\ 
        \rowcolor{rowgray}
        \cite{43_Tee2014_webKM} Tee et al. \textit{Web-based knowledge management model for managing and sharing green knowledge of software development in ...} 
        (2014) \\ 
        \cite{27_Wedyan2023_Testing} Wedyan et al. \textit{Integration and unit testing of software energy consumption} (2023) \\ 
        \rowcolor{rowgray}
        \cite{40_Wei2023_GreenCodeGeneration} Wei et al. \textit{Towards greener yet powerful code generation via quantization: an empirical study} (2023) \\ 
        \cite{41_Zhang2023_NetAI} Zhang et al. \textit{Toward net-zero carbon emission in network AI for 6G and beyond} (2023) \\
        \hline    
    \end{tabular}
\end{table}

The list of candidate studies was completed by running backward snowballing \cite{Wohlin2014_GuidelinesSnowballing}, which resulted in 30 additional candidate studies.
These were again classified for inclusion/exclusion by repeating the same process as described above and by directly considering their full read. 
As a result, 22 additional studies were selected, reaching a total of 65 studies to be considered by our SLR.
The final list of selected studies is shown in \Cref{tab:design:selected-studies}.

\subsection{Literature Classification}
\label{sec:design:classification}
The selected studies were classified by exploiting the taxonomy in \Cref{tab:design:taxonomy}. 
The taxonomy is rooted in the 5W1H questions driving our research (\Cref{sec:design:rq}), with each of these questions being mapped to one or more possible classifications, \ie categories of values.
\rev{For the sake of brevity, \Cref{tab:design:taxonomy} lists for each category the values associated with at least one selected study.}

\begin{table}[h]
    \caption{Taxonomy used to classify the selected studies in relation to our 5W1H questions. \enquote{*} marks the categories for which multiple different of their values can be associated with a selected study.}
    \label{tab:design:taxonomy}
    \footnotesize
\begin{tabular}{lllp{.47\textwidth}}
    \hline
    \rowcolor{headergray}
    \textbf{5W1H} & \multicolumn{2}{l}{\textbf{Category}} & \textbf{Values} \\  
    \hline 
    \textbf{Who?} & \multicolumn{2}{l}{stakeholder*} & \it end user, IT operator, product owner, software architect, software provider, software developer, sustainability engineer  \\
    \rowcolor{rowgray}
    \textbf{What?} & proposal & sustainability guidelines & \it experience report, good practices \\
    \rowcolor{rowgray}
    & & reference models & \it reference model \\
    \rowcolor{rowgray}
    & & measurement solutions & \it  assistive tooling, estimate technique \\
    \rowcolor{rowgray}
    & & software improvement techniques & \it  decision making, distributed protocol, ML, model-driven, optimisation, search-based software engineering, static analysis, testing  \\
    \rowcolor{rowgray}
    & \multicolumn{2}{l}{source code availability} & \it black-box, white-box \\
    \rowcolor{rowgray}
    & \multicolumn{2}{l}{TRL} & \it 1--9 \\
    \textbf{Why?} & goal* & \rev{indirectly} & \it measure, reduce, optimise \\
    & & \rev{directly} & \it measure, reduce, optimise \\
    \rowcolor{rowgray}
    \textbf{When?} & \multicolumn{2}{l}{lifecycle stage*} & \it requirement analysis, design, implementation, integration \& testing, rollout \\
    \textbf{Where?} & \multicolumn{2}{l}{target application domain} & \it blockchain, cloud computing, computer vision, cyber-physical systems, edge computing, generic software, HPC, ML, mobile applications \\
    \rowcolor{rowgray}
    \textbf{How?} & \multicolumn{2}{l}{publication year} & \it 2023, 2022, 2021, ... \\
    \rowcolor{rowgray}
    & \multicolumn{2}{l}{publication type} & \it book chapter, conference paper, journal article \\
    \hline
\end{tabular}
\end{table}

\smallskip \noindent 
\textbf{Who?}
The taxonomy enables classifying the (potentially multiple) \textit{stakeholders} for the selected studies. 
These include 
the \textit{end-user} (who selects and uses a software product),
the \textit{IT operator} (who operates a software product),
the \textit{product owner} (who manages the software process and aims at maximising the value of a software product),
the \textit{software architect} (who designs of a software product),
the \textit{software developer} (who implements parts of a software product),
the \textit{software provider} (who delivers a software product to end-users), and 
the \textit{sustainability engineer} (who engineers the sustainability of a software product).

\smallskip \noindent 
\textbf{What?}
The taxonomy in \Cref{tab:design:taxonomy} enables classifying the \textit{proposal} of each selected study by distinguishing among \textit{sustainability guidelines} (in the form of shared \textit{good practices} or \textit{experience report}s), \textit{reference models} (namely, abstract frameworks consisting of interlinked concepts aimed at clarifying what is \rev{carbon-efficiency} for software design and development), \textit{measurement solutions} (in the form of \textit{assistive tooling} or \textit{estimate techniques} supporting the measurement of a software's sustainability), and \textit{software improvement techniques} (namely, techniques aimed at supporting the design and/or development of software to \rev{reduce its carbon footprint}). 

The taxonomy also enables distinguishing whether the proposal of a selected study is \textit{white-box} (or \textit{black-box}), namely whether it requires the availabilty of a software's source code, \rev{given its relevance to software design and development}. 
It is also possible to indicate the \textit{TRL} (Technology Readiness Level) of a selected study's proposal from 1 to 9, following the TRL classification defined by the European Commission~\cite{EC2014_TRL}. 

\smallskip \noindent 
\textbf{Why?}
The taxonomy in \Cref{tab:design:taxonomy} enables classifying the (potentially multiple) \textit{goals} of the selected studies. 
Namely, it enables indicating whether a selected study aims at supporting the design and development of software while \rev{\textit{measuring}, \textit{reducing}, or \textit{optimising} its carbon emissions}.
\rev{While reducing and optimising carbon emissions will certainly improve a software's carbon efficiency, measurement can also support this, for example by comparing a software's carbon footprint before and after updates. This is to check whether such updates have made the software more or less sustainable. We also distinguish whether this is done by \textit{directly} measuring a software's carbon emissions, or \textit{indirectly}, by assuming that reducing a software's energy consumption will also reduce its carbon footprint.}

\smallskip \noindent 
\textbf{When?}
The taxonomy in \Cref{tab:design:taxonomy} enables distinguishing the (potentially multiple) \textit{lifecycle stage}s supported by a selected study's proposal. 
To this end, we follow the iterative software lifecycle process described by Kneuper \cite{Kneuper2018_SDLC}, which consists of \textit{requirement analysis}, \textit{design}, \textit{implementation}, \textit{integration \& testing}, and \textit{rollout}, in that order.

\smallskip \noindent 
\textbf{Where?}
The taxonomy in \Cref{tab:design:taxonomy} makes it possible to distinguish the \textit{target application domain} for a selected study.
These include \textit{blockchain}s, \textit{cloud computing}, \textit{computer vision}, \textit{cyber-physical systems}, \textit{edge computing}, \textit{HPC}, \textit{ML}, and \textit{mobile applications}.
The above listed domains are complemented by the \textit{generic software} value, which is used when the proposal is intended to work with any software, rather than targeting and exploiting the peculiarities of a specific application domain.

\smallskip \noindent 
\textbf{How?}
The taxonomy in \Cref{tab:design:taxonomy} enables classifying eliciting the publication trends for the selected studies.
More precisely, it enables indicating the \textit{publication year} for each selected study, as well as its \textit{publication type} (\ie \textit{book chapter}, \textit{conference paper}, or \textit{journal article}).

\subsection{Replication Package}
\label{sec:design:replication-package}
To enable repeating our study, as well as verifying our findings, we publicly released a replication package on Zenodo \cite{Danushi2024_ReplicationPackage}.
The replication package includes the intermediate artifacts produced during the search and selection of studies to be considered, as well as their final classification. 
The selected studies can instead be accessed by relying on the bibliographic information provided by the references available in \Cref{tab:design:selected-studies}.

\section{Overviewing the State-of-the-Art Solutions}
\label{sec:sota}
This section overviews the whole corpus of our SLR, which constitutes the state of the art in the field of designing and developing environmentally sustainable software. Particularly, we hereinafter present the surveyed articles by dividing them as per the \textit{What?} classification from \Cref{sec:design:classification}, namely sustainability guidelines (\Cref{sec:sota:guidelines}), reference models (\Cref{sec:sota:refmodels}), measurement solutions (\Cref{sec:sota:measurement}), and software improvement techniques (\Cref{sec:sota:sitechs}).
To support the reader, the classification of the selected studies is graphically recapped in \Cref{tab:classification}.

\begin{table}[p]
    \caption{Classification of the selected studies. For the sake of conciseness, we use some abbreviations for proposals (\ie \textit{AT=assistive tooling, DM=decision making, DP=distributed protocol, ER=experience report, ET=estimate technique, GP=good practices, MD=model-driven, OP=optimisation, RM=reference model, SA=static analysis, SB=search-based software engineering, and TE=testing}), source code availability (\textit{BB=black-box and WB=white-box}), and publication type (\textit{BC=book chapter, CP=conference paper, and JA=journal article}).} 
    \label{tab:classification}
    \centering
    \resizebox{.98\textwidth}{!}{
\begin{tabular}{|c|l|l|l|l|l|c|l|c|c|c|c|l|c|l|l|c|l|c|c|c|c|c|c|c|c|}
\hline
\multirow{2}{*}{} & \multicolumn{7}{c|}{\multirow{2}{*}{\textbf{Who?}}} & \multicolumn{4}{c|}{\multirow{2}{*}{\textbf{What?}}} & \multicolumn{6}{c|}{\textbf{Why?}} & \multirow{2}{*}{\textbf{Where?}} & \multicolumn{5}{c|}{\multirow{2}{*}{\textbf{When?}}} & \multicolumn{2}{c|}{\multirow{2}{*}{\textbf{How?}}} \\ \cline{13-18}
 & \multicolumn{7}{c|}{} & \multicolumn{4}{c|}{} & \multicolumn{3}{c|}{\rev{indirectly}} & \multicolumn{3}{c|}{\rev{directly}} & & \multicolumn{5}{c|}{} & \multicolumn{2}{c|}{} \\ \cline{2-26} 
\textbf{Ref.} & {\rotatebox{90}{end user}} & {\rotatebox{90}{IT operator}} & {\rotatebox{90}{product owner}} & {\rotatebox{90}{sust. engineer}} & {\rotatebox{90}{software architect}} & {\rotatebox{90}{software developer \,}} & {\rotatebox{90}{software provider}} & \multicolumn{2}{c|}{\rotatebox{90}{proposal}} & {\rotatebox{90}{white-/black-box}} & \rotatebox{90}{TRL} & {\rotatebox{90}{measure}} & {\rotatebox{90}{reduce}} & {\rotatebox{90}{minimise}} & {\rotatebox{90}{measure}} & {\rotatebox{90}{reduce}} & {\rotatebox{90}{minimise}} & \rotatebox{90}{\centering target app. domain} & {\rotatebox{90}{requirement analysis}} & {\rotatebox{90}{design}} & {\rotatebox{90}{development}} & {\rotatebox{90}{integration \& testing \,}} & \rotatebox{90}{roll-out} & {\rotatebox{90}{publication type}} & \rotatebox{90}{publication year} \\ \hline
\cite{16_Ahmed2020_codingCloud} & & & & & & \checkmark & \checkmark & \cellcolor{guidelines} sustainability guidelines & \cellcolor{guidelines} {GP} & {WB} & 2 & & \checkmark & & & \checkmark & & cloud computing & & \checkmark & \checkmark & & & {JA} & 2020 \\ \hline
\cite{31_Alofi2023_Self-Optimizing} & & & & \checkmark & \checkmark & & & \cellcolor{techniques} sw improvement techniques & \cellcolor{techniques} {SB} & {BB} & 4 & & & \checkmark & & & \checkmark & blockchain & & & \checkmark & & & {JA} & 2023 \\ \hline
\cite{30_Alofi2022_OptimizingConsensus} & & & & \checkmark & \checkmark & & & \cellcolor{techniques} sw improvement techniques & \cellcolor{techniques} {SB} & {BB} & 4 & & & \checkmark & & & \checkmark & blockchain & & & \checkmark & & & {JA} & 2022 \\ \hline
\cite{18_Atkinson2014_GreenSpecifications} & \checkmark & & \checkmark & & & \checkmark & \checkmark & \cellcolor{refmodels} reference models & \cellcolor{refmodels} {RM} & {BB} & 3 & & & & & \checkmark & & cloud computing & \checkmark & \checkmark & & & & {JA} & 2014 \\ \hline
\cite{50_Becker2016_Requirements} & & & \checkmark & \checkmark & & \checkmark & \checkmark & \cellcolor{refmodels} reference models & \cellcolor{refmodels} {RM} & {BB} & 2 & & \checkmark & & & \checkmark & & generic sw & \checkmark & \checkmark & \checkmark & & & {JA} & 2016 \\ \hline
\cite{21_Beghoura2014_GreenEvaluation} & & & & & & \checkmark & & \cellcolor{techniques} sw improvement techniques & \cellcolor{techniques} {ML} & {BB} & 4 & \checkmark & & & & & & mobile app. & \checkmark & & & & & {CP} & 2014 \\ \hline
\cite{17_Brownlee2021_AccuracyEnergyML} & & & & & & \checkmark & & \cellcolor{guidelines} sustainability guidelines & \cellcolor{guidelines} {ER} & {BB} & 4 & & \checkmark & & & & & ML & & \checkmark & \checkmark & & & {CP} & 2021 \\ \hline
\cite{01_Chauhan2013_GreenSDLCCloud} & \checkmark & & & & & \checkmark & \checkmark & \cellcolor{guidelines} sustainability guidelines & \cellcolor{guidelines} {GP} & {BB} & 1 & & \checkmark & & & \checkmark & & cloud computing & \checkmark & \checkmark & \checkmark & \checkmark & \checkmark & {JA} & 2013 \\ \hline
\cite{36_Fernandez2018_GreenDesign} & & & & & \checkmark & & & \cellcolor{guidelines} sustainability guidelines & \cellcolor{guidelines} {GP} & {BB} & 2 & & \checkmark & & & \checkmark & & generic sw & \checkmark & \checkmark & & & & {CP} & 2018 \\ \hline
\cite{24_Dick2013_GSE-Agile} & \checkmark & & & \checkmark & & \checkmark & & \cellcolor{refmodels} reference models & \cellcolor{refmodels} {RM} & {BB} & 2 & & \checkmark & & & \checkmark & & generic sw & \checkmark & \checkmark & \checkmark & \checkmark & \checkmark & {CP} & 2013 \\ \hline
\cite{48_Dick2010_GreenSustainableSoftware} & \checkmark & \checkmark & & & \checkmark & \checkmark & & \cellcolor{refmodels} reference models & \cellcolor{refmodels} {RM} & {BB} & 2 & & \checkmark & & & \checkmark & & generic sw & \checkmark & \checkmark & \checkmark & \checkmark & \checkmark & {BC} & 2010 \\ \hline
\cite{32_Fakhar2012_GreenComputing} & & & & & & \checkmark & & \cellcolor{techniques} sw improvement techniques & \cellcolor{techniques} {SA} & {WB} & 3 & & \checkmark & & & \checkmark & & cloud computing & \checkmark & \checkmark & \checkmark & \checkmark & \checkmark & {JA} & 2012 \\ \hline
\cite{52_Fu2021_ComputerVision} & & & & & & \checkmark & & \cellcolor{guidelines} sustainability guidelines & \cellcolor{guidelines} {GP} & {BB} & 3 & & \checkmark & & & \checkmark & & comp. vis. & & \checkmark & & & & {CP} & 2021 \\ \hline
\cite{51_LeGoaer2021_greenCode} & & & & & & \checkmark & & \cellcolor{techniques} sw improvement techniques & \cellcolor{techniques} {SA} & {WB} & 3 & & \checkmark & & & & & mobile app. & & \checkmark & \checkmark & & & {CP} & 2021 \\ \hline
\cite{13_LeGoaer2023_ecoCodeProject} & & & & & & \checkmark & & \cellcolor{techniques} sw improvement techniques & \cellcolor{techniques} {SA} & {WB} & 3 & & \checkmark & & & \checkmark & & mobile app. & & & \checkmark & & & {CP} & 2023 \\ \hline
\cite{49_LeGoaer2023_ecoCode} & & & & & & \checkmark & & \cellcolor{techniques} sw improvement techniques & \cellcolor{techniques} {SA} & {WB} & 4 & & \checkmark & & & \checkmark & & mobile app. & & & \checkmark & & & {CP} & 2023 \\ \hline
\cite{15_Heithoff2023_DigitalTwins} & & & \checkmark & & & \checkmark & & \cellcolor{techniques} sw improvement techniques & \cellcolor{techniques} {MD} & {BB} & 1 & \checkmark & & & \checkmark & & & cyb. phys. sys. & \checkmark & \checkmark & \checkmark & \checkmark & \checkmark & {CP} & 2023 \\ \hline
\cite{39_Hoesch-Klohe2010_GreenBPM} & \checkmark & & & & \checkmark & \checkmark & & \cellcolor{techniques} sw improvement techniques & \cellcolor{techniques} {MD} & {BB} & 4 & & \checkmark & & & & & generic sw & & \checkmark & & & & {CP} & 2010 \\ \hline
\cite{34_Ibrahim2021_GSDProcessModel} & \checkmark & & & & & \checkmark & & \cellcolor{refmodels} reference models & \cellcolor{refmodels} {RM} & {BB} & 2 & & \checkmark & & & \checkmark & & generic sw & & \checkmark & & & & {JA} & 2021 \\ \hline
\cite{25_Ibrahim2022_GreenFactors} & \checkmark & & & & & \checkmark & & \cellcolor{refmodels} reference models & \cellcolor{refmodels} {RM} & {BB} & 3 & & \checkmark & & & \checkmark & & generic sw & \checkmark & \checkmark & \checkmark & \checkmark & \checkmark & {JA} & 2022 \\ \hline
\cite{07_Jayanthi20211_OrganizationalStructure} & \checkmark & & & \checkmark & & \checkmark & & \cellcolor{refmodels} reference models & \cellcolor{refmodels} {RM} & {BB} & 1 & & \checkmark & & & \checkmark & & generic sw & \checkmark & \checkmark & \checkmark & \checkmark & \checkmark & {CP} & 2021 \\ \hline
\cite{26_Jebraeil2017_gUML} & & & & & \checkmark & & & \cellcolor{techniques} sw improvement techniques & \cellcolor{techniques} {MD} & {BB} & 4 & & \checkmark & & & \checkmark & & cloud computing & & \checkmark & & & & {CP} & 2017 \\ \hline
\cite{46_Johann2011_IntegratedApproach} & \checkmark & \checkmark & & & \checkmark & \checkmark & & \cellcolor{refmodels} reference models & \cellcolor{refmodels} {RM} & {BB} & 2 & & \checkmark & & & \checkmark & & generic sw & \checkmark & \checkmark & \checkmark & \checkmark & \checkmark & {CP} & 2011 \\ \hline
\cite{03_Karunakaran2013_PetriNet} & & & \checkmark & & & & & \cellcolor{techniques} sw improvement techniques & \cellcolor{techniques} {MD} & {BB} & 4 & & & & & \checkmark & & generic sw & \checkmark & \checkmark & \checkmark & \checkmark & \checkmark & {CP} & 2013 \\ \hline
\cite{54_Kern2013_GreenSA} & \checkmark & \checkmark & & & & \checkmark & & \cellcolor{refmodels} reference models & \cellcolor{refmodels} {RM} & {BB} & 2 & & \checkmark & & & & & generic sw & & & \checkmark & & & {CP} & 2013 \\ \hline
\cite{61_Kern2015_ICTCO2footprint} & & & \checkmark & & & \checkmark & & \cellcolor{measurement} measurement solutions & \cellcolor{measurement} {ET} & {BB} & 4 & & & & \checkmark & & & generic sw & \checkmark & \checkmark & \checkmark & \checkmark & \checkmark & {BC} & 2015 \\ \hline
\cite{33_Kern2018_AssessCriteria} & \checkmark & & \checkmark & \checkmark & & \checkmark & & \cellcolor{refmodels} reference models & \cellcolor{refmodels} {RM} & {BB} & 3 & & \checkmark & & & & & generic sw & & \checkmark & & & & {JA} & 2018 \\ \hline
\cite{64_Kharchenko2013_VonNeumannParadigm} & & & & \checkmark & & & & \cellcolor{guidelines} sustainability guidelines & \cellcolor{guidelines} {GP} & {BB} & 2 & \checkmark & & & & & & generic sw & \checkmark & \checkmark & \checkmark & \checkmark & \checkmark & {CP} & 2013 \\ \hline
\cite{62_Kipp2011_GreenMetrics} & & & & \checkmark & & & & \cellcolor{measurement} measurement solutions & \cellcolor{measurement} {ET} & {BB} & 2 & & \checkmark & & & & & HPC & \checkmark & \checkmark & \checkmark & \checkmark & \checkmark & {CP} & 2011 \\ \hline
\cite{28_Kipp2012_GreenIndicators} & & & & \checkmark & & & & \cellcolor{measurement} measurement solutions & \cellcolor{measurement} {ET} & {BB} & 3 & & & & & \checkmark & & HPC & \checkmark & \checkmark & \checkmark & \checkmark & \checkmark & {JA} & 2012 \\ \hline
\cite{65_Kocak2019_UtilityModel} & \checkmark & & & & & & \checkmark & \cellcolor{techniques} sw improvement techniques & \cellcolor{techniques} {DM} & {BB} & 2 & & \checkmark & & & & & generic sw & \checkmark & \checkmark & & & & {CP} & 2013 \\ \hline
\cite{57_Lago2019_DecisionMaps} & & & & & \checkmark & & & \cellcolor{techniques} sw improvement techniques & \cellcolor{techniques} {DM} & {BB} & 3 & & \checkmark & & & \checkmark & & generic sw & & \checkmark & & & & {CP} & 2019 \\ \hline
\cite{11_Lago2011_SOSE} & \checkmark & & & & & \checkmark & & \cellcolor{refmodels} reference models & \cellcolor{refmodels} {RM} & {BB} & 2 & & \checkmark & & & \checkmark & & generic sw & \checkmark & \checkmark & \checkmark & & & {CP} & 2011 \\ \hline
\cite{44_Lami2012_SoftwareMeasurement} & & & \checkmark & & & & & \cellcolor{refmodels} reference models & \cellcolor{refmodels} {RM} & {BB} & 2 & & \checkmark & & & \checkmark & & generic sw & \checkmark & \checkmark & \checkmark & \checkmark & \checkmark & {CP} & 2012 \\ \hline
\cite{06_Lami2014_MeasurementFramework} & & & \checkmark & & & & & \cellcolor{refmodels} reference models & \cellcolor{refmodels} {RM} & {BB} & 2 & & & & & \checkmark & & generic sw & \checkmark & \checkmark & \checkmark & \checkmark & \checkmark & {CP} & 2014 \\ \hline
\cite{02_Lami2013_SustainabilityIndicators} & & & \checkmark & & & & & \cellcolor{refmodels} reference models & \cellcolor{refmodels} {RM} & {BB} & 2 & & & & & \checkmark & & generic sw & \checkmark & \checkmark & \checkmark & \checkmark & \checkmark & {CP} & 2013 \\ \hline
\cite{59_Lasla2022_GreenPoWAE} & & & & & & \checkmark & & \cellcolor{techniques} sw improvement techniques & \cellcolor{techniques} {DP} & {BB} & 4 & & \checkmark & & & & & blockchain & & & \checkmark & & & {JA} & 2022 \\ \hline
\cite{58_Mahmoud2013_GreenModel} & & & & & & \checkmark & & \cellcolor{refmodels} reference models & \cellcolor{refmodels} {RM} & {WB} & 2 & & \checkmark & & & \checkmark & & generic sw & \checkmark & \checkmark & \checkmark & \checkmark & \checkmark & {JA} & 2013 \\ \hline
\cite{19_Mancebo2021_FEETINGS} & \checkmark & \checkmark & & & & \checkmark & & \cellcolor{techniques} sw improvement techniques & \cellcolor{techniques} {TE} & {WB} & 4 & & \checkmark & & & & & generic sw & & & & \checkmark & & {BC} & 2021 \\ \hline
\cite{08_Mehra2023_AssessRefactoring} & & & & & & \checkmark & & \cellcolor{guidelines} sustainability guidelines & \cellcolor{guidelines} {ER} & {WB} & 5 & & \checkmark & & & \checkmark & & generic sw & & & \checkmark & \checkmark & \checkmark & {CP} & 2023 \\ \hline
\cite{38_Sharma2022_GreenQuotient} & & & & & & \checkmark & & \cellcolor{refmodels} reference models & \cellcolor{refmodels} {RM} & {BB} & 3 & & \checkmark & & & & & generic sw & \checkmark & \checkmark & \checkmark & \checkmark & \checkmark & {CP} & 2015 \\ \hline
\cite{56_Mohankumar2016_GSDLCM} & & & & & & \checkmark & & \cellcolor{measurement} measurement solutions & \cellcolor{measurement} {AT} & {BB} & 2 & & \checkmark & & & & & generic sw & \checkmark & \checkmark & \checkmark & \checkmark & \checkmark & {CP} & 2022 \\ \hline
\cite{05_Moshnyaga2013_LCA} & & & & & & \checkmark & & \cellcolor{refmodels} reference models & \cellcolor{refmodels} {RM} & {BB} & 2 & & \checkmark & & & & & generic sw & \checkmark & \checkmark & \checkmark & \checkmark & \checkmark & {JA} & 2016 \\ \hline
\cite{09_Moshnyaga2013_AssessLContribution} & & & \checkmark & & & & & \cellcolor{measurement} measurement solutions & \cellcolor{measurement} {ET} & {BB} & 4 & \checkmark & & & \checkmark & & & generic sw & \checkmark & \checkmark & \checkmark & \checkmark & \checkmark & {CP} & 2013 \\ \hline
\cite{29_Moshnyaga2013_AssessLC} & & & \checkmark & & & & & \cellcolor{measurement} measurement solutions & \cellcolor{measurement} {ET} & {BB} & 4 & \checkmark & & & \checkmark & & & generic sw & \checkmark & \checkmark & \checkmark & \checkmark & \checkmark & {CP} & 2013 \\ \hline
\cite{45_Naumann2011_GREENSOFT} & & & \checkmark & & & \checkmark & & \cellcolor{measurement} measurement solutions & \cellcolor{measurement} {ET} & {BB} & 4 & \checkmark & & & \checkmark & & & mobile app. & \checkmark & \checkmark & \checkmark & \checkmark & \checkmark & {CP} & 2013 \\ \hline
\cite{60_Naumann2015_QualityModels} & \checkmark & \checkmark & & & \checkmark & \checkmark & & \cellcolor{refmodels} reference models & \cellcolor{refmodels} {RM} & {BB} & 2 & & \checkmark & & & \checkmark & & generic sw & \checkmark & \checkmark & \checkmark & \checkmark & \checkmark & {JA} & 2001 \\ \hline
\cite{04_Noureddine2012_GREENS} & \checkmark & & & \checkmark & & \checkmark & & \cellcolor{refmodels} reference models & \cellcolor{refmodels} {RM} & {BB} & 2 & & \checkmark & & & \checkmark & & generic sw & \checkmark & \checkmark & \checkmark & \checkmark & \checkmark & {BC} & 2015 \\ \hline
\cite{12_Pa2017_GreenDesignMeasurement} & & \checkmark & & & & \checkmark & & \cellcolor{measurement} measurement solutions & \cellcolor{measurement} {ET} & {BB} & 4 & \checkmark & & & & & & generic sw & & & \checkmark & & \checkmark & {CP} & 2012 \\ \hline
\cite{14_Pan2022_CNNDecomposition} & & & & & & \checkmark & & \cellcolor{measurement} measurement solutions & \cellcolor{measurement} {AT} & {BB} & 3 & & \checkmark & & & \checkmark & & generic sw & \checkmark & \checkmark & \checkmark & \checkmark & \checkmark & {CP} & 2017 \\ \hline
\cite{53_Patterson2022_MLCO2footprint} & & & & & \checkmark & & & \cellcolor{guidelines} sustainability guidelines & \cellcolor{guidelines} {GP} & {BB} & 4 & & & & & \checkmark & & ML & & & \checkmark & & & {CP} & 2022 \\ \hline
\cite{63_Penzenstadler2013_GenericModel} & & & & & & \checkmark & & \cellcolor{guidelines} sustainability guidelines & \cellcolor{guidelines} {GP} & {BB} & 4 & & \checkmark & & & \checkmark & & ML & & & \checkmark & & & {JA} & 2022 \\ \hline
\cite{10_Ponsard2018_CaseStudy} & & & & \checkmark & \checkmark & & & \cellcolor{refmodels} reference models & \cellcolor{refmodels} {RM} & {BB} & 3 & & \checkmark & & & \checkmark & & generic sw & \checkmark & & & & & {CP} & 2013 \\ \hline
\cite{22_Radersma2022_GreenEvaluation} & & & & & \checkmark & & & \cellcolor{guidelines} sustainability guidelines & \cellcolor{guidelines} {ER} & {BB} & 3 & & \checkmark & & & & & generic sw & \checkmark & \checkmark & \checkmark & & & {CP} & 2018 \\ \hline
\cite{35_Raisian2021_MeasurementStructure} & & & & & & \checkmark & & \cellcolor{guidelines} sustainability guidelines & \cellcolor{guidelines} {GP} & {BB} & 2 & & \checkmark & & & & & HPC & & & \checkmark & & & {JA} & 2022 \\ \hline
\cite{47_Sharma2015_GreenSDLC} & & & & & & \checkmark & & \cellcolor{refmodels} reference models & \cellcolor{refmodels} {RM} & {BB} & 2 & \checkmark & & & \checkmark & & & generic sw & & & \checkmark & & & {CP} & 2021 \\ \hline
\cite{55_Shenoy2011_GreenDevelopmentModel} & \checkmark & & & & & \checkmark & & \cellcolor{guidelines} sustainability guidelines & \cellcolor{guidelines} {GP} & {BB} & 3 & & \checkmark & & & \checkmark & & generic sw & & & \checkmark & & & {JA} & 2022 \\ \hline
\cite{23_Siegmund2022_GreenConfiguration} & & & \checkmark & & & \checkmark & & \cellcolor{measurement} measurement solutions & \cellcolor{measurement} {AT} & {BB} & 4 & & \checkmark & & & \checkmark & & ML & \checkmark & \checkmark & \checkmark & & & {CP} & 2023 \\ \hline
\cite{20_Sikanda2023_GreenAIQuotient} & & & & & & \checkmark & & \cellcolor{guidelines} sustainability guidelines & \cellcolor{guidelines} {GP} & {BB} & 2 & & \checkmark & & & \checkmark & & generic sw & \checkmark & \checkmark & \checkmark & \checkmark & \checkmark & {CP} & 2011 \\ \hline
\cite{42_Simon2023_ImpactSLC} & & & \checkmark & & & & & \cellcolor{measurement} measurement solutions & \cellcolor{measurement} {ET} & {BB} & 4 & \checkmark & & & \checkmark & & & generic sw & \checkmark & \checkmark & \checkmark & \checkmark & \checkmark & {CP} & 2023 \\ \hline
\cite{37_Tee2014_GreenKnowledge} & & \checkmark & & & & \checkmark & & \cellcolor{refmodels} reference models & \cellcolor{refmodels} {RM} & {BB} & 2 & & \checkmark & & & \checkmark & & generic sw & \checkmark & \checkmark & \checkmark & \checkmark & \checkmark & {CP} & 2014 \\ \hline
\cite{43_Tee2014_webKM} & & \checkmark & & & & \checkmark & & \cellcolor{refmodels} reference models & \cellcolor{refmodels} {RM} & {BB} & 2 & & \checkmark & & & \checkmark & & generic sw & \checkmark & \checkmark & \checkmark & \checkmark & \checkmark & {CP} & 2014 \\ \hline
\cite{27_Wedyan2023_Testing} & & & & & & \checkmark & & \cellcolor{techniques} sw improvement techniques & \cellcolor{techniques} {TE} & {WB} & 2 & \checkmark & & & & & & generic sw & & & & \checkmark & & {CP} & 2023 \\ \hline
\cite{40_Wei2023_GreenCodeGeneration} & & & & & & \checkmark & & \cellcolor{techniques} sw improvement techniques & \cellcolor{techniques} {ML} & {BB} & 4 & & \checkmark & & & \checkmark & & ML & & \checkmark & \checkmark & & & {CP} & 2023 \\ \hline
\cite{41_Zhang2023_NetAI} & & & & & & & \checkmark & \cellcolor{techniques} sw improvement techniques & \cellcolor{techniques} {OP} & {BB} & 4 & & & \checkmark & & & \checkmark & edge computing & \checkmark & \checkmark & \checkmark & \checkmark & \checkmark & {JA} & 2023 \\ \hline
\end{tabular}
}
\end{table}

\subsection{Sustainability Guidelines}
\label{sec:sota:guidelines}

In this section, we focus on the portion of the surveyed corpus that proposes sustainability-related guidelines to design and develop sustainable software. 
We start by illustrating those studies that focus on \rev{indirectly reducing the carbon footprint} of running software. We then move to \rev{those directly considering the carbon footprint of software}, and finally discuss those works that \rev{follow a hybrid approach}. We organise our discussion by target application domain.

\smallskip \noindent \textbf{\rev{Indirect}}. For {generic software}, Kharchenko and Gorbenko~\cite{64_Kharchenko2013_VonNeumannParadigm} provide a set of {good practices} in the form of procedural steps intended for sustainability engineers to implement green management of system components. Such guidelines account for different aspects of system components (e.g., threats, fault-tolerance) and enable identifying existing pathological and evolutionary chains among those aspects. From the analysis of such chains it is possible to determine how to minimise the system's resource (and energy) consumption, while improving a set of customisable quality metrics (e.g., reliability, availability, security, resilience). 

On a different line, addressing generic software architects, Ponsard et al.~\cite{10_Ponsard2018_CaseStudy} conduct an {experience report} of an industrial case study about building and putting in operation a shared pick-up service. They show how to conduct the requirement analysis phase by also considering sustainability goals and their intertwining with the functional requirements set by other stakeholders of a software system. By distinguishing between green \textit{by} software and green \textit{in} software requirements, they take implementation and architectural decisions at different levels (e.g., choice of efficient programming language and algorithms, use of lightweight virtualisation) and define key performance indicators to monitor for improving future releases. 

Moving to approaches that consider {HPC} as their application domain, Radersma~\cite{22_Radersma2022_GreenEvaluation} lists a set of {good practices} for software developers to implement more environmentally sustainable and maintainable software artefacts, e.g., by adopting green programming languages, avoiding unnecessary data accesses in favour or computing results on-the-fly, employing state-of-the-art HPC hardware, and relying on cluster power management software to optimise energy consumption. Considering instead an {experience report} on {ML}, Brownlee et al.~\cite{17_Brownlee2021_AccuracyEnergyML} rely on exhaustive and genetic search to (sub-)optimally tune the hyperparameters of multi-layer perceptron networks. From a software developer's perspective, they investigate the relations among different configurations, accuracy and energy consumption in both the training and inference phases on a set of benchmark networks. As a result, in their benchmark, they reduce energy consumption by 30--50\% with a decrease in accuracy below 1\%.

\smallskip \noindent \textbf{\rev{Direct}}. Still in the ML domain, Pan and Rajan \cite{14_Pan2022_CNNDecomposition} are the sole to discuss {good practices} for {software architects} to directly reduce {carbon emissions} produced by convolutional neural networks. Instead of training convolutional neural networks from scratch, they recommend exploiting model decomposition -- having each module capable of classifying a single output class -- together with transfer, one-shot, and few-shot learning to favour reusability and replaceability and (consequently) reduce the carbon emissions associated with model (re)training.


\smallskip \noindent \textbf{\rev{Hybrid}}. 
In collaboration with an industry partner and targeting software developers, Fernandez and Lago~\cite{36_Fernandez2018_GreenDesign} identify a set of domain generic green strategies and link them to the prioritisation of software quality requirements (e.g. usability, maintainability). Their aim is threefold: to raise people's awareness of software sustainability, to make service operations environmentally sustainable, and to support sustainable development processes, by emphasising the need to consider sustainability throughout the software design phase and beyond. Similarly, Shenoy and Eeratta \cite{55_Shenoy2011_GreenDevelopmentModel} suggest that quality and sustainability should be considered equally important throughout the whole software lifecycle. They propose a set of {improvement practices} for {software developers} to drive green software development.

Still targeting {generic software}, Mehra et al. \cite{08_Mehra2023_AssessRefactoring} report on their experience of using a static analysis tool to identify code inefficiencies and act upon them suitably. 
Their experience report, clearly intended for application developers, showcases the potential of static analysis in detecting energy-inefficient patterns, assessing their impact, and providing remediation strategies.
They indeed successfully refactor a large application codebase from an industrial use case.
Siegmund et al. \cite{23_Siegmund2022_GreenConfiguration} analyse the relationship between software configuration and energy consumption. They propose ML-based code analysis methods to identify energy and carbon inefficiencies in software, from the developer's perspective. On the other hand, they enable software users to select energy-optimised configurations for their software according to a black-box model. Their model accounts for non-monotonic energy consumption in the configuration space, which enables relying non-linear functions to model the system's behaviour.

Building upon the work by Shenoy and Eeratta \cite{55_Shenoy2011_GreenDevelopmentModel}, Chauhan and Saxena \cite{01_Chauhan2013_GreenSDLCCloud} propose the \textit{Green Cloud Framework} (GCF) as a set of {best practices} to handle software development lifecycle in cloud computing settings. Their work accounts for recommendations from organisations like Greenpeace to use clean and renewable energy sources in software development to make software development energy-aware and carbon-aware. Specifically, GCF is intended for software providers, developers and end-users, to make Service Level Agreements (SLAs) include guarantees of sustainability metrics (i.e. carbon emissions and energy consumption).
Differently, \cite{55_Shenoy2011_GreenDevelopmentModel} mainly focuses on software developers, without envisioning the key role of SLAs between application providers and end-users. 

Ahmed \cite{16_Ahmed2020_codingCloud} introduces a {white-box} technique for software developers to study cloud software via direct access to the source code. He proposes a new step to optimise the software development lifecycle, named parameterized development phase, between the design and implementation phases, useful to define parameters related to the used programming language and hardware architecture, and their energy consumption. This new step is useful to identify margins for improvement as early as possible in the design phase to allow for code tuning in a cost-effective manner.

Speaking of ML, Patterson et al. \cite{53_Patterson2022_MLCO2footprint} discuss the energy consumption and carbon footprint of machine learning training and inference, also considering embodied carbon emissions. They eventually propose the so-called 4M best practices for software developers, which involve $(i)$ choosing greener energy locations for ML workloads ({Map}), $(ii)$ using the latest possible hardware ({Machine}), $(iii)$ relying on cloud resources to reduce the environmental footprint ({Mechanisaton}), and $(iv)$ exploit state of the art lightweight neural networks ({Models}). Similarly, Fu et al. \cite{52_Fu2021_ComputerVision} provide {good practices} for {software developers} to compute carbon emissions costs for the search phase of a computer vision architecture based on deep learning over its entire lifecycle. 

\subsection{Reference models}
\label{sec:sota:refmodels}

In this section, we focus on the largest part of our surveyed corpus, which proposes reference models for the design and development of carbon-aware applications. As in the previous section, we first report on contributions that \rev{indirectly} reduce carbon emissions, and then move to those that focus on \rev{direct} carbon reduction. Finally, we discuss \rev{hybrid} proposals.

\smallskip \noindent \textbf{\rev{Indirect}}.
%
Sharma et al. \cite{47_Sharma2015_GreenSDLC} and Mohankumar and Kumar \cite{56_Mohankumar2016_GSDLCM} introduce a {reference model} for {software developers}, by relying and extending the software lifecycle model of~\cite{Kneuper2018_SDLC}. Sharma et al. \cite{47_Sharma2015_GreenSDLC} consider an iterative software development lifecycle made of six stages, \ie requirement analysis, green analysis, design, implementation, testing, and maintenance. The green analysis phase cyclically connects to all others and acts as a checkpoint entity to monitor and continuously enhance the sustainability of all other phases according to set guidelines. Also following an iterative approach, Mohankumar and Kumar \cite{56_Mohankumar2016_GSDLCM} only consider a four-stage lifecycle consisting of requirements analysis, design, coding and testing. For the requirements analysis, they propose to $(i)$ use questions, guidelines, and generic templates to match customers' needs with sustainable technology choices, $(ii)$ use well-timed metrics data for the purpose, and $(iii)$ formal methods to avoid ambiguities. Based on requirements specification, developers should write the documentation concerning green and sustainable templates for designing all application components. Then, throughout the implementation phase they should always adopt best practices, e.g. avoid unnecessary classes, closing database connections after use, avoid stand-alone installation in favour of on-demand remote access.

The reference model by Kern et al.~\cite{54_Kern2013_GreenSA} goes towards defining an eco-labelling to drive {software developers} in the design and development phases. It can be applied by analysing which tasks or functions of the software are most frequently used and then designing case studies accordingly, also considering servers' energy consumption. This way of proceeding allows developers and end-users to compare and contrast different software implementations by determining the most sustainable one. Subsequent proposals by Kern et al. are also aimed at {IT operators}~\cite{54_Kern2013_GreenSA} and for {sustainability engineers} and {project owners}~\cite{33_Kern2018_AssessCriteria}.

\smallskip \noindent \textbf{\rev{Direct}}. To reduce carbon footprint and to raise green and sustainable awareness in generic software, Lami et al. \cite{06_Lami2014_MeasurementFramework, 02_Lami2013_SustainabilityIndicators} define a {methodological approach} for {product owners}. They use such an approach to derive sustainability indicators for the whole lifecycle of a software project and use them to populate bi-dimensional matrices, known as hierarchical software templates. Such templates are used to explore three levels of environmental effects, viz. direct, indirect, and systemic, also associating direct effects with their measurements and success factors. Last, in compliance with ISO/IEC 33000, Lami et al. \cite{06_Lami2014_MeasurementFramework} define the assessment of six process sustainability levels, viz. incomplete, performed, managed, defined, measured, and optimised. 

Considering instead the peculiarities of cloud applications, Atkinson et al. \cite{18_Atkinson2014_GreenSpecifications} propose a {reference model} that is intended for several stakeholders such as {product owners}, {software providers}, {developers}, and {end-users}. The model focuses on the specification of SLAs to integrate energy costs and environmental impact (\eg usage of renewable energy sources) among the considered QoS parameters. They propose adopting different UML modelling to better communicate with all stakeholders by relying on structural, functional, and behavioural views. With an eye to economical sustainability, their proposal also models different pricing policies of the software offer.

\smallskip \noindent \textbf{\rev{Hybrid}}. All studies on \rev{hybrid} reference models are intended for the generic software domain with a black-box approach.
For example, Tee et al. \cite{37_Tee2014_GreenKnowledge,43_Tee2014_webKM} define a {reference model} for both {software developers} and {IT operators}, by proposing a procedural method for making software development lifecycle more sustainable. They aim at gathering, assembling and storing of the information for the knowledge management of a green and sustainable software development lifecycle. They collect information on existing findings, processes and activities, and techniques in software development, and propose a knowledge management model and step-by-step guidelines to make software sustainable. They then map techniques to actionable tasks to minimise power consumption and carbon footprint while developing software. Following up on the above, Tee et al.~\cite{37_Tee2014_GreenKnowledge} extend the knowledge-management model into a green-management model to be shared among all stakeholders to build an action-oriented and collaborative community of practice.

Jayanthi et al.~\cite{07_Jayanthi20211_OrganizationalStructure} and Dick et al.~\cite{24_Dick2013_GSE-Agile} provide a {reference model} for {software developers}, {sustainability engineers}, and {end-users} by defining green and sustainable software development lifecycle steps integrated within Agile operations. In particular, Jayanthi et al. \cite{07_Jayanthi20211_OrganizationalStructure} argue that organisations should establish a sustainability department that includes quality assurance engineers for the requirements analysis phase, user experience designers for the design phase, certified green computing developers for the coding and testing phase, and managers for the overall success of the sustainable software deployment. Conversely, Dick et al. \cite{24_Dick2013_GSE-Agile} add a specific sustainable development stage to the software lifecycle model. They associate SCRUM activities to their driving role among sustainability engineers and software developers, and embed sustainability reporting and retrospective in the Agile development model.

On a similar line, extending their previous effort~\cite{06_Lami2014_MeasurementFramework}, Lami and Buglione \cite{44_Lami2012_SoftwareMeasurement} provide a {reference model} for {project owners} to enact measurements by considering a cause-effect diagram analysis on the infrastructure, people, software process and product. The model has to create a suitable measurement plan that integrates all needed information elements to specifically address the sustainability improvement throughout software development. They add sustainability to project management as a first-class citizen like quality, time, and budget.
With an akin objective, Ibrahim et al. \cite{34_Ibrahim2021_GSDProcessModel,25_Ibrahim2022_GreenFactors} provide a {reference model} for both {software developers} and {end-users}. They focus holistically on software product development, as well as the related hardware disposal and recycling. Their approach also relies on protocols for (semi-)structured interviews with involved stakeholders to collect data useful for process mining from a sustainability viewpoint. The proposed model aims at achieving resource and energy efficiency to pursue long-term improvement of software sustainability, with a particular focus on reducing waste (\eg over-requirements, miscommunication, rework, and lack of knowledge). Ibrahim et al.  \cite{34_Ibrahim2021_GSDProcessModel} also showcase the proposed approach over real industrial use cases.

Raisian et al. \cite{35_Raisian2021_MeasurementStructure} propose green and sustainable software measurements as a {reference model} for {software developers} that aim at targeting sustainability by considering the environmental dimension (\ie energy and carbon footprint in processing and storage), the economic dimension (\ie usability, available time and costs), and the social dimension (\ie support tools, human well-being, performance). To achieve the above, they aim at identifying suitable hierarchical measurements that are to be evaluated, documented, and assessed for continuous software enhancement.
Still with a hierarchical approach to continuous software improvement, Penzenstadler and Femmer~\cite{63_Penzenstadler2013_GenericModel} provide a {reference model} definition based on different levels of granularity, for both {software architects} and {sustainability engineers}. After defining sustainability metrics and their targets, the proposed model foresees matching software lifecycle activities to their (positive/negative) impacts on energy or carbon emissions, and identifying responsible personnel for each activity. 

Becker et al. \cite{50_Becker2016_Requirements} propose a {reference model} that allows {product owners, software providers, developers} and {sustainability engineers} to compare two equivalent software systems designed and implemented with and without considering sustainability, respectively.
The model aligns the sustainability goals with those of the system, defines the project scope and system boundaries for any stakeholder, makes the risk analysis for each sustainability dimension, and finally, implements requirements and specifications based on templates including checklists for sustainability criteria and standards compliance in all the interrelated dimensions.
Lago and Jansen \cite{11_Lago2011_SOSE} propose a {reference model} that supports the iterative and incremental way of operating for {software developers} and {end-users}. The model is used to drive the implementation of two Web services, \ie the Environmental Strategies as-a-Service and the Green Metrics as-a-Service, useful to make the environmental impact of software services visible by continuously monitoring it, giving feedback and tailoring software based on the available data to reduce energy and carbon footprints.

We conclude by presenting the  GREENSOFT reference model, originally defined by Naumann et al.~\cite{45_Naumann2011_GREENSOFT}, which led to many other research efforts. Following a holistic approach, GREENSOFT is intended for all stakeholders, \ie {software architects}, {developers}, {IT operators}, and {end-users}. The model seconds the Agile software development model and considers first (green-IT), second (green-by-IT) and third order (systemic) effects related to designing, implementing, using and decommissioning software products. It foresees the definition of {ad hoc} sustainability criteria and metrics, the analysis of process models for the development, administration, purchase and use of software, and the adoption of recommendations and tools for all stakeholders. Agile sprints are enhanced by including sustainability reviews and retrospectives that focus on measured metrics and criteria, monitored via a sustainability journal throughout requirements elicitation, design, implementation and testing. These steps are mentioned by Dick et al. \cite{24_Dick2013_GSE-Agile} when considering the sustainable Agile development stage.

Retaking the process assessment model of GREENSOFT,  Johann et al. \cite{46_Johann2011_IntegratedApproach} propose a {reference model} to develop sustainable software products, while mitigating negative impacts and amplifying positive impacts based on a knowledge base shared among different stakeholders. Their proposal includes a first implementation of recommendations and tools in the form of checklists and suggestions. Dick et al.~\cite{48_Dick2010_GreenSustainableSoftware} also extend GREENSOFT by relying on the definition of checklists and best practices realised through stakeholder interviews, with a focus on software development and project management. Particularly, they highlight how sustainability decisions taken at a set step of the software lifecycle only show their impact during the next one.

Later, Mahmoud and Ahmad \cite{58_Mahmoud2013_GreenModel} re-consider the GREENSOFT model and discuss how software itself can be used as a tool to aid and promote fine-grained green computing for operating systems, frameworks, performance monitoring, programs written for energy management purposes, and virtualisation. Their model also accounts for fault-tolerance and failure management.
Finally, following GREENSOFT and the ISO/IEC 12207 standard,\footnote{Systems and software engineering — Software lifecycle processes, International Organization for Standardization, 2008} Naumann et al.~\cite{60_Naumann2015_QualityModels} define criteria to identify software quality aspects as per a process-centric model, inherent both to the final product and the implemented process, through lifecycle assessment. Their model addresses {software developers, sustainability engineers} and {end-users} and is given in a full-fledged Agile version following the whole lifecycle as well as in a lightweight version focussing on process assessment and sustainability reviews and previews.

\subsection{Measurement Solutions}
\label{sec:sota:measurement}

This section focuses on the portion of the analysed corpus about measurement solutions for software sustainability, which take the form of {estimate techniques} and {assistive tooling}. Their aim is to measure or estimate carbon emissions related to software design and development so as to make informed choice throughout the software lifecycle.

\smallskip \noindent \textbf{\rev{Indirect}}. Starting from those approaches that \rev{indirectly consider carbon emissions}, Noureddine et al.~\cite{04_Noureddine2012_GREENS} develop a real-time software energy monitoring framework intended for both {software developers} and {end-users}. Such solution reports estimates on energy consumption by relying on a service-oriented architecture consisting of a monitoring sensor, a database, and a lifecycle monitoring module. Specifically, the individual components are isolated and managed using start/stop options, thus providing flexibility while continuously monitoring the deployment hardware (\ie processors, network interfaces). The approach is illustrated over different programs written in different programming languages.

Focussing instead on the HPC domain, Kipp et al. \cite{62_Kipp2011_GreenMetrics} introduce an {estimate technique} intended for {sustainability engineers}. They propose Green Performance Indicators as a supplement to Key Performance Indicators encompassing factors such as the use of virtual/real IT resources, the service lifecycle, approximate environmental impact, and organisational considerations. They consequently assess sustainability at different levels of a system architecture - from the abstract view to the business logic and up to the entire IT service centre. 
The methodology involves monitoring clusters of metrics for energy-intensive applications by first analysing single jobs and observing the relation between energy consumption and execution time for each run. Such data is then input to a constraint-based formulation to optimise resource usage and energy consumption. 

\smallskip \noindent \textbf{\rev{Direct}}. Sharma et al. \cite{38_Sharma2022_GreenQuotient} devise an {assistive tooling} for {software developers} to assess the carbon efficiency of their projects throughout the development lifecycle. This tool provides a standard set of metrics to evaluate the energy consumption impact of different design decisions at each stage of the project. The approach is based on a series of questions related to project characteristics, \eg programming languages, used third-party libraries, cloud providers, technologies, processes, and teams' organisation. Ultimately, the tool provides an overall sustainability score and recommendations to further improve software sustainability. 
To also promote responsibility, awareness and transparency, Kern et al. \cite{61_Kern2015_ICTCO2footprint} develop a model- and tool-based calculation method intended for {software developers} and {product owners} to measure the carbon footprint of their products according to international standards. The strategy involves collecting data throughout the software development lifecycle to continuously monitor carbon emissions.

\smallskip \noindent \textbf{\rev{Hybrid}}. This last section contains a large number of surveyed studies, scattered across a plethora of different domains. Starting from generic software, Pa et al.~\cite{12_Pa2017_GreenDesignMeasurement} propose an {assistive tool} to help developers in understanding and implementing sustainable software by estimating its carbon footprint, also based on its energy consumption. The prototype \rev{features} online accessible web-based dashboards to visualise sustainability reports and suggestions for defining best trade-off solutions between database organisation, network usage and human resources. The tool can also track historical data to monitor project evolution over time. 
Simon et al. \cite{42_Simon2023_ImpactSLC} provide a comprehensive {estimate technique} based on graph modelling regarding the environmental impact of software lifecycle for {project owners}. They have nodes representing lifecycle stages and edges describing different environmental impacts based on direct observations. Lifecycle assessment follows a lightweight bottom-up approach to identify stakeholders' responsibility and actionable insights and to identify hot spots for potential improvement. Still targeting project owners, Moshnyaga~\cite{05_Moshnyaga2013_LCA, 09_Moshnyaga2013_AssessLContribution} analyses the total energy consumption associated with the software throughout its entire development lifecycle and evaluates its contribution to carbon emissions, also accounting for hardware-specific emissions. The proposed methodology also relies on rapid prototyping to build software following an iterative three-step process consisting of business model definition, risk analysis (revision), and development plan (update). The same authors also tailored their proposal specifically targeting mobile applications~\cite{29_Moshnyaga2013_AssessLC}, along with both {project owners} and {software developers}.

Speaking of the ML domain, based on Mehra et al.~\cite{38_Sharma2022_GreenQuotient},  Sikand et al. \cite{20_Sikanda2023_GreenAIQuotient} present a standard set of metrics for measuring the energy efficiency of artificial intelligence systems throughout the development lifecycle. They also call attention to harmful aspects associated with {red AI}, \eg the exponential growth in data, model complexity, large models, increased number of experiments, and the relentless pursuit of higher performance. Extending their previous effort~ \cite{62_Kipp2011_GreenMetrics}, Kipp et al. \cite{28_Kipp2012_GreenIndicators} instead focus on HPC by introducing \textit{Green Performance Indicators} to complement the traditional KPIs, by enabling a comprehensive evaluation of the environmental impact and sustainability classification of IT service centres by examining various hierarchical system levels, from the application level to the middle-ware and up to the infrastructure.

\subsection{Software Improvement Techniques}
\label{sec:sota:sitechs}

We last explore the portion of our surveyed corpus that proposes specific improvement techniques to design and develop sustainable software, \eg  {decision-making}, {model-driven}, {static analysis}, {testing}, {optimisation}, {search-based software engineering}, and {distributed protocols}.

\smallskip \noindent \textbf{\rev{Indirect}}. Kocak and Alptekin \cite{65_Kocak2019_UtilityModel} propose a multi-criteria {decision-making technique} for generic software, as part of the {analysis} and {design} of carbon efficient software. Their technique aims at balancing between the needs of {end-users}, who want energy-efficient software at a low cost, and {software providers}, who offer products at a maximum quality level. The proposal addresses the conflicting interests by determining demand and supply functions for decision-making entities that are considered rational, intelligent, and having a comprehensive plan of action for any situation that may arise. The proposal also discusses the implementation of a strategic interaction based on the interdependencies of software product properties and attributes. They also align their proposal with the ISO/IEC 25010 standards, including attributes such as functional suitability, reliability, performance efficiency, and usability. The authors then propose a method to translate these requirements into terms of price, quality, and utility. 

To ensure software functionality compliance and measure energy consumption, Wedyan et al.~\cite{27_Wedyan2023_Testing} propose a {white-box analysis technique} for {testing} purposes, to {software developers}. They advocate the use of static analysis to construct call graphs for each class, with methods denoted by nodes, and control between methods denoted by edges. They then associate an estimate of energy consumption per computation unit and recommend considering cyclomatic complexity, which is a measure of software complexity based on the number of decision points and lines of code involved. Their testing framework ultimately reports both the energy consumed and the success of the test itself. 
Similarly, Mancebo et al. \cite{19_Mancebo2021_FEETINGS} propose a {white-box analysis technique} for {testing} purposes to {software developers, IT operators} and {end-users}. Their prototype framework is capable of gathering energy consumption data during software execution, as well as of processing them to generate statistics and reports for comparison and analysis. They provide users with a conceptual framework for green and sustainable software measurement, a methodological guide to conduct energy consumption measurements, and a prototype tester.

Focussing on mobile applications, Le Goa\"{e}r \cite{51_LeGoaer2021_greenCode} provides a code scanning tool for the Android Studio IDE. The tool performs {white-box static analyses} to identify green code smells and associated potential energy optimisation.
The scanning tool is designed for {software developers} who use the Android Lint framework. Although individual energy savings may be small for single Android devices, the cumulative impact across billions of devices is significant. Different green bugs are matched to their severity, priority, and category, along with candidate single-click fixes (\eg rely on dark UI, data compression, timeout wave locks).
Still in the mobile landscape, Beghoura et al. \cite{21_Beghoura2014_GreenEvaluation} instead propose a {black box} analysis for {software developers} that relies on a supervised vector machine model to create specific energy consumption models for different types of devices. They investigate the energy consumption patterns of hardware when running specific applications, accounting for various factors such as hardware configuration, product design, and battery ageing. More specifically, the study aims at understanding non-linear energy consumption components, such as changes in CPU power consumption as a function of frequency and voltage. 

Moving to the field of blockchains, Lasla et al. \cite{59_Lasla2022_GreenPoWAE} introduce a {distributed protocol}, that defines a green proof-of-work algorithm. The implementation comprises two rounds. The first involves all network nodes in the mining process, while the second involves a smaller set of nodes, called runners-up. The energy consumption is reduced since the same energy spent to mine one block is also used to select a small number of miners to exclusively mine the next block. The protocol also considers and balances among usual blockchain tradeoffs such as fork frequency, security and decentralisation level.

\smallskip \noindent \textbf{\rev{Direct}}. Only the proposal by Karunakaran and Rao \cite{03_Karunakaran2013_PetriNet}, intended for project owners, provides a {model-driven technique} that solely considers the carbon footprint of software. They propose to exploit coloured Petri Nets that embed time constraints and hierarchical implementation to devise and configure optimal policies of a waterfall software development lifecycle. Their model accounts for machine, software and resource levels to simulate programming, testing and debugging of software projects and estimate the associated carbon emissions.

\smallskip \noindent \textbf{\rev{Hybrid}}.
Hoesch-Klohe et al. \cite{39_Hoesch-Klohe2010_GreenBPM} provide a {model-driven technique} which leverages business process models for their capability to capture the relations between resources and activities, specifically focusing on carbon emission impact, through textual assertions. Their technique involves identifying all process designs, ordering them based on their abstraction level, and then systematically determining a suitable association to computing resources accounting both for carbon emissions and energy consumption.
Lago~\cite{57_Lago2019_DecisionMaps} devises a {decision-making technique} intended for {software architects}, based on  architectural designs and assessments for sustainability. In the specific, decision maps capture the impacts of sustainability dimensions as immediate, enabling, and systemic, and categorise their effects as positive, negative, or undecided. Decision maps are used to explore the software design space by addressing multiple sustainability perspectives at the same time; they can be used to drive the software design decision process, while refining it iteratively.

Focussing on cyber-physical systems, Heithoff et al. \cite{15_Heithoff2023_DigitalTwins} instead present a model-based approach for {project owners} and {software developers}. By relying on digital twins and digital shadow systems, they monitor the sustainability of the actual real-world software. Digital twins are designed to be smaller in size than the original system and allow for faster analyses through simulated behaviour.  Their prototype also enables monitoring energy consumption peaks in comparison to current workloads and predefined levels. 
With cloud computing developers as their target, Jebraeil et al.~\cite{26_Jebraeil2017_gUML} propose a {model-driven technique} called g-UML. They extend UML diagrams with contextual information about data centre and network operations, \ie power generation, cooling and servers consumption, data communication footprint. This enables the efficient identification of energy-related issues and the organisation of power-related components into an architectural view. Resulting gUML diagrams can assist software developers in writing energy-efficient software, enabling collaboration between hardware and software architects.
Addressing software developers, Fakhar et al. \cite{32_Fakhar2012_GreenComputing} combine a {white-box static analysis technique} for energy-efficient software compilation with a {black-box analysis technique} for effective task scheduling. The compiler translates source code in an energy-efficient format, optimises binary code through green and sustainable strategies, and provides hints for energy-efficient coding. On the other hand, the scheduler utilizes a greedy algorithm to efficiently make use of cloud resources, selecting and allocating machines based on user requirements while aiming at using as many processing cores in a machine as possible.

Moving from cloud to edge computing settings, Zhang et al. \cite{41_Zhang2023_NetAI} present an {optimisation technique} tailored for {software providers} operating in a network that offers smart services. They propose the use of an evaluation framework that combines dynamic energy trading with task allocation optimisation for federated edge intelligence networks to reduce carbon emissions throughout the software lifecycle. At design time, they mainly focus on computing carbon emissions related to data-generating devices. At development time, they account for emissions due to pre-training, testing, parameter selection and optimisation, based on specified service requirements and usage scenarios. At deployment time, they address model distribution, while, at disposal time, they focus on generalisation, modification, and transfer issues.

In line with Le Goa\"{e}r~\cite{51_LeGoaer2021_greenCode}, Le Goa\"{e}r et al.~\cite{13_LeGoaer2023_ecoCodeProject, 49_LeGoaer2023_ecoCode} present a {white-box static}, computer-aided, customisable code inspection tool for Android {software developers}, named ecoCode. This tool is designed to identify and refactor code structures that are inefficient in terms of energy consumption (thus impacting on carbon emissions). By emphasising energy efficiency as a quality attribute, ecoCode aims at promoting the concept of green and sustainable code, akin to the way security and maintainability are enforced in clean code. Particularly, ecoCode performs a scan of project repositories to identify environmental issues, which are then displayed on a web dashboard.

To reduce re-training costs in ML, Wei et al. \cite{40_Wei2023_GreenCodeGeneration} propose a {machine learning technique} for {software developers}. Specifically, they suggest using model compression through a quantization method for the tasks of code generation (from natural language to code), summarisation (the counterpart, from code to natural language), and type inference. This technique involves representing model parameters with fewer bits, making the model smaller in size so as to increase the speed of operations. Empirical results validate the effectiveness of this approach on multiple code-specific tasks, demonstrating that it is green and sustainable while preserving accuracy and robustness compared to the full precision counterpart model.

Finally, in the domain of blockchains, Alofi et al. \cite{31_Alofi2023_Self-Optimizing, 30_Alofi2022_OptimizingConsensus} propose a {search-based software engineering technique} for addressing energy and carbon consumption concerns. The proposal is intended for {software architects} and {sustainability engineers} and employs evolutionary meta-heuristics to identify subsets of miners that maximise trust levels in the system, maintain decentralisation and reduce energy consumption and carbon emissions. They propose to implement the system as an autonomic MAPE-K loop that ensures enactment of taken decisions.

\section{Answering our 5W1H Questions}
\label{sec:5w1h}
This section aims at answering to our 5W1H questions on the design and development of environmentally sustainable software (\Cref{sec:design:rq}).
More precisely, by zooming out and considering the overall classification given by \Cref{tab:classification}, we aim at identifying trends and coverage on \textit{who} are the targeted stakeholders (\Cref{sec:5W1H:who}), \textit{what} methods are proposed to design and develop environmentally sustainable software (\Cref{sec:5W1H:what}), \textit{why} (\Cref{sec:5W1H:why}), \textit{where} and \textit{when} the proposed solutions are applied (\Cref{sec:5W1H:where,sec:5W1H:when}, respectively), and \textit{how} the selected studies have been presented to the scientific community (\Cref{sec:5W1H:how}).

\subsection{Who?}
\label{sec:5W1H:who}
\begin{figure}
    \centering
    \includegraphics[width=.92\textwidth,trim={0 1cm 0 1cm}]{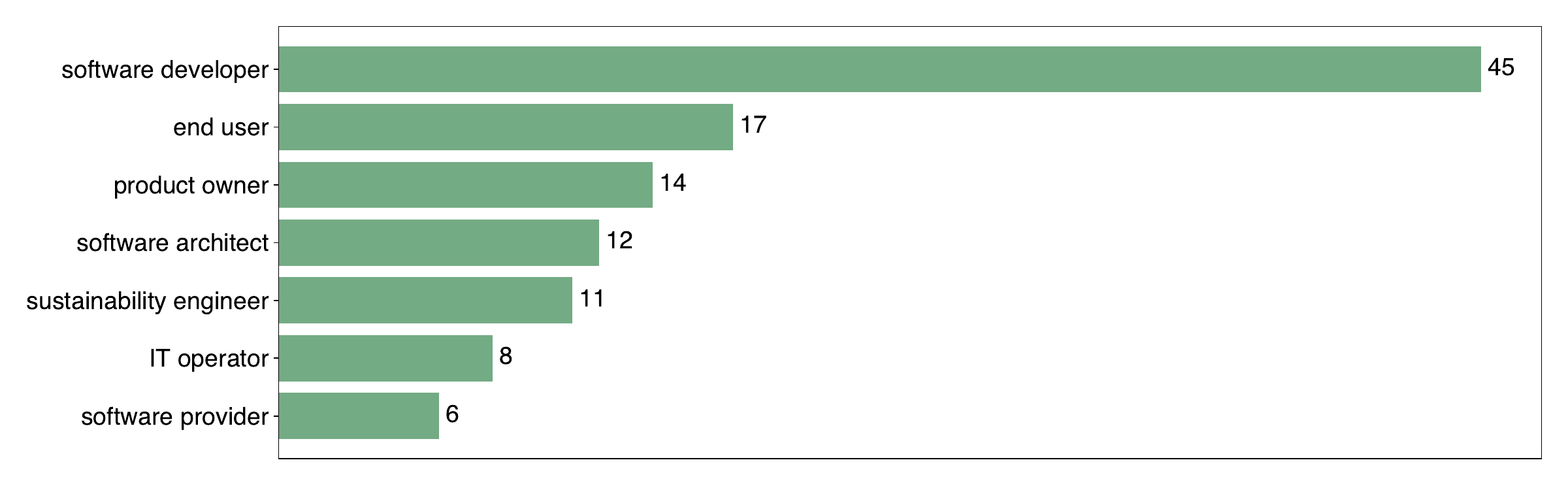}
    \caption{Frequency of stakeholders over the selected studies. Since some studies target multiple stakeholders, the sum of frequencies is higher than the number of selected studies.}
    \label{fig:5w1h:who-stakeholders}
\end{figure}

The \textit{who} of the design and development of environmentally sustainable software aims at identifying who are the stakeholders targeted by the state of the art solutions available for the topic.
\Cref{fig:5w1h:who-stakeholders} lists the stakeholders considered by the selected studies, as well as their frequencies.\footnote{In this and the following subsections, we refer to the frequency of a given classification in the selected studies, measured by the number of studies classified as such.}

From \Cref{fig:5w1h:who-stakeholders}, there is an evident unbalance between design- and development-related stakeholders.
Indeed, \textit{software developer}s are by far the most frequent stakeholders, \rev{who are} targeted by the vast majority of selected studies (\ie 45 out of the 65 selected studies).
The frequency of \textit{software developer}s is also significantly higher than that of software design-related stakeholders (\ie \textit{software architect}s and \textit{sustainability engineer}s), even when added together.
Given that our focus is on the design \textit{and} development of sustainable software, we would have expected a more balanced distribution among design- and development-related stakeholders. 
In this perspective, and to avoid putting all the burden of environmental sustainability on \textit{software developer}s, there is a need for more solutions targeting \textit{software architect}s and \textit{sustainability engineer}s, who should be better supported in the design of environmentally sustainable software.

Other interesting observations come from the coverage of stakeholders not directly involved in software design and development, like \textit{end user}s,  \textit{product owner}s, \textit{IT operator}s, and \textit{software provider}s.
While the frequency of \textit{IT operator}s may be justified by the emergent wave of DevOps practices \cite{Azad2023_DevOpsSLR}, the frequencies of the other stakeholders seem to suggest that the design and development of environmentally sustainable software naturally involves more stakeholders than just those involved in the design and development of software. 
This holds especially for \textit{end user}s and \textit{product owner}s, whose frequencies are almost the same as those of design-related stakeholders, since they can concretely impact on the design and development of environmentally sustainable software.
Indeed, environmental sustainability can be a discriminant for the software chosen by \textit{end user}s \cite{07_Jayanthi20211_OrganizationalStructure,11_Lago2011_SOSE,18_Atkinson2014_GreenSpecifications,19_Mancebo2021_FEETINGS}.
Additionally, the choices taken by \textit{product owner}s along the software process can affect the environmental sustainability of the software itself \cite{06_Lami2014_MeasurementFramework,05_Moshnyaga2013_LCA,20_Sikanda2023_GreenAIQuotient,29_Moshnyaga2013_AssessLC}. 
The above are just first insights on how different stakeholders impact on the design and development of environmentally sustainable software, even if not directly participating in software design and development themselves.
Such insights shed light on the need for further investigation on the roles, contributions, and support needed by all the possible stakeholders for the design and development of environmentally sustainable software.

\begin{highlights}
    \highlight{1}{\textit{Software developer}s are by far the most frequent stakeholders for the existing solutions to support the design and development of environmentally sustainable software}
    \highlight{2}{More support should be provided to \textit{software architect}s and \textit{sustainability engineer}s for the design of environmentally sustainable software}
    \highlight{3}{The design and development of environmentally sustainable software involves multiple different stakeholders, whose roles, contributions, and needed support should be further investigated}
\end{highlights}

\subsection{What?}
\label{sec:5W1H:what}
We hereby consider \textit{what} are the state of the art solutions available for designing and developing environmentally sustainable software.
More precisely, we aim at providing an overall view of the available proposals, whether they directly inspect source code, and their overall TRL, based on the classification of selected studies provided in \Cref{tab:classification}.

\begin{figure}
    \centering
    \includegraphics[width=.92\textwidth,trim={0 1cm 0 1cm}]{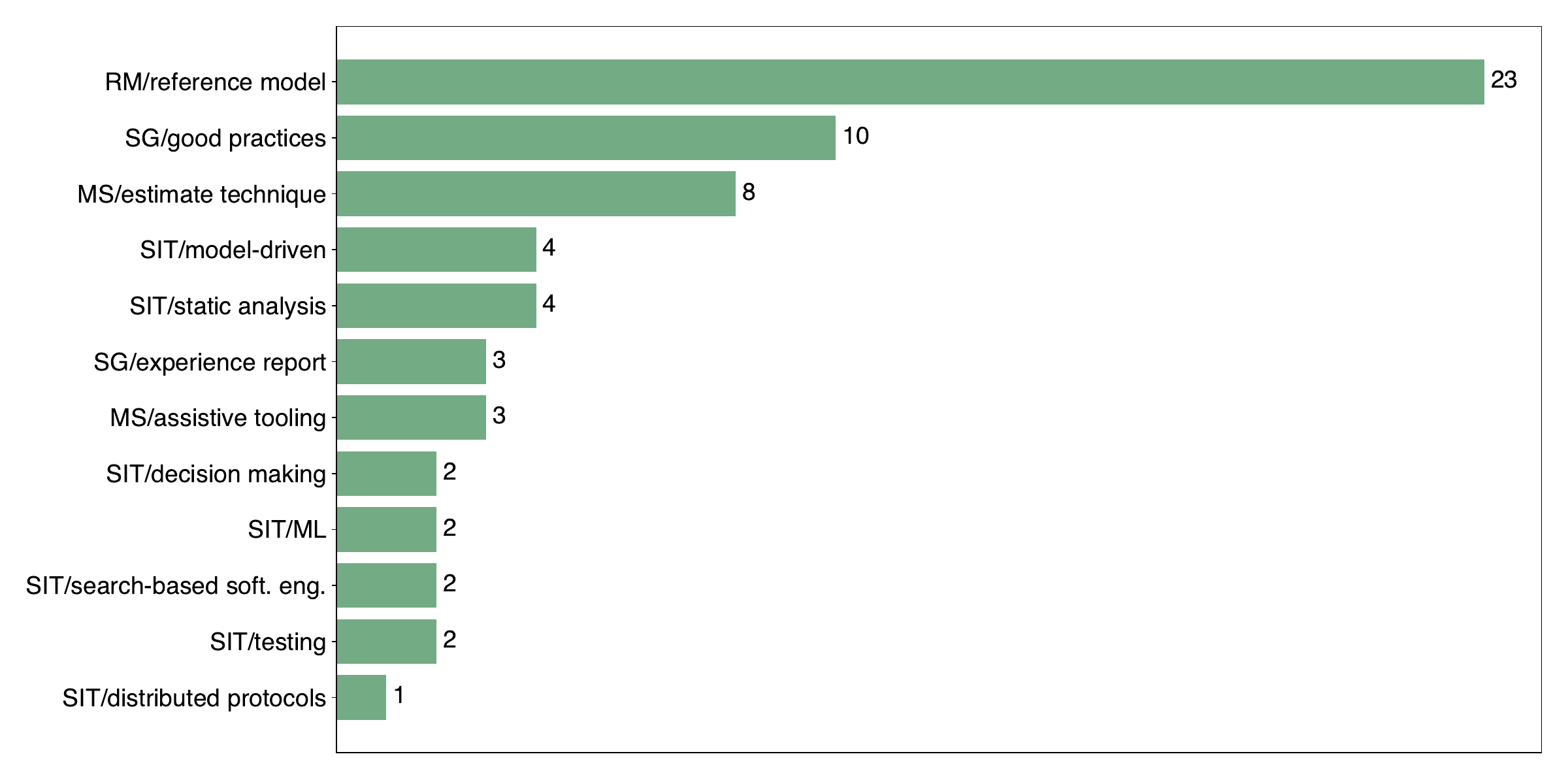}
    \caption{Frequency of proposals in the selected studies (with \textit{MS=measurement solutions}, \textit{RM=reference models}, \textit{SG=sustainability guidelines}, and \textit{SIT=software improvement techniques}).}
    \label{fig:5w1h:what-proposal}
\end{figure}

\Cref{fig:5w1h:what-proposal} displays the frequency of proposals in the selected studies, clearly outlining a predominance of \textit{reference model}s, proposed in 23 out of the 65 selected studies.
Of these, six studies focus on GREENSOFT, a reference model first proposed by Naumann et al. \cite{45_Naumann2011_GREENSOFT} and then rivisited and extended in five subsequent studies \cite{24_Dick2013_GSE-Agile,46_Johann2011_IntegratedApproach,48_Dick2010_GreenSustainableSoftware,58_Mahmoud2013_GreenModel,60_Naumann2015_QualityModels}.
The remaining efforts propose different reference models, however still sharing some commonalities with GREENSOFT and with each other.
Whilst the number of reference models witnesses the need for clarifying what is environmental sustainability for software design and development, the lack of a commonly accepted reference model calls for standardisation efforts along this line.

The existing proposals for designing and developing environmentally sustainable software then mainly focus on providing a first support to the purpose, namely sharing guidelines in the form of \textit{good practices} or providing \textit{estimate techniques} to assess the environmental sustainability of the software under design/development.
The rest is then scattered among different attempts (especially, \textit{software improvement techniques}) to support the design and development of environmentally sustainable software, however without any concrete proposal emerging as the most promising or used for the task.
Further research is, therefore, needed to establish which techniques can effectively support designing and developing environmentally sustainable software. 

\begin{figure}
    \centering
    \begin{minipage}{.49\textwidth}
        \centering \footnotesize
        \includegraphics[width=.92\textwidth,trim={0 1cm 0 1cm}]{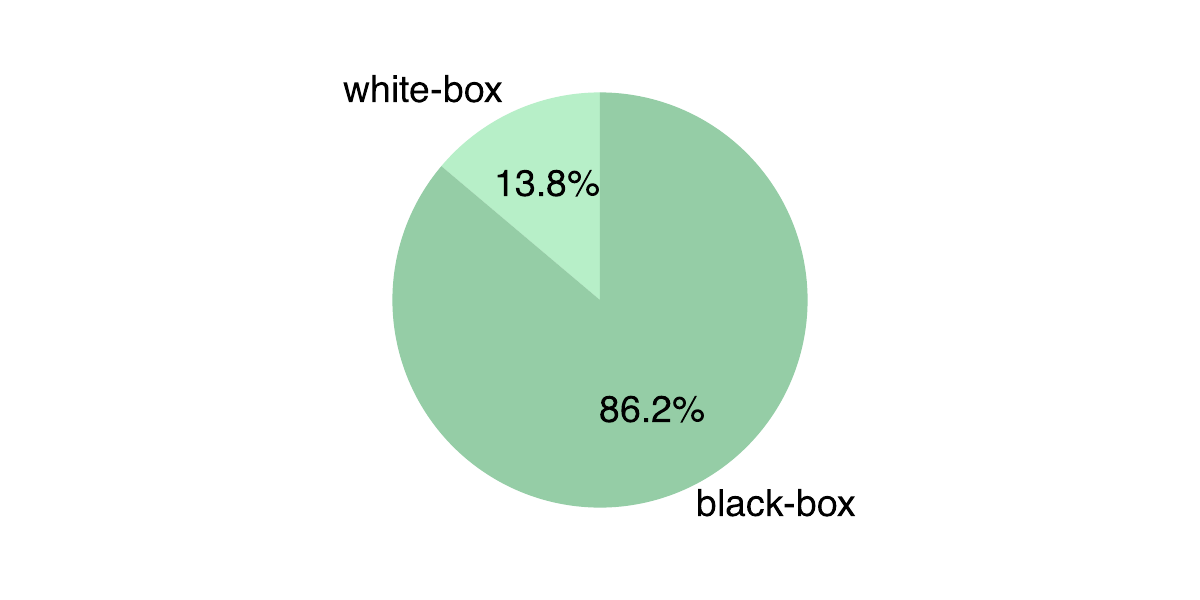}
        (a)
    \end{minipage}
    \begin{minipage}{.49\textwidth}
        \centering \footnotesize
        \includegraphics[width=.92\textwidth,trim={0 1cm 0 1cm}]{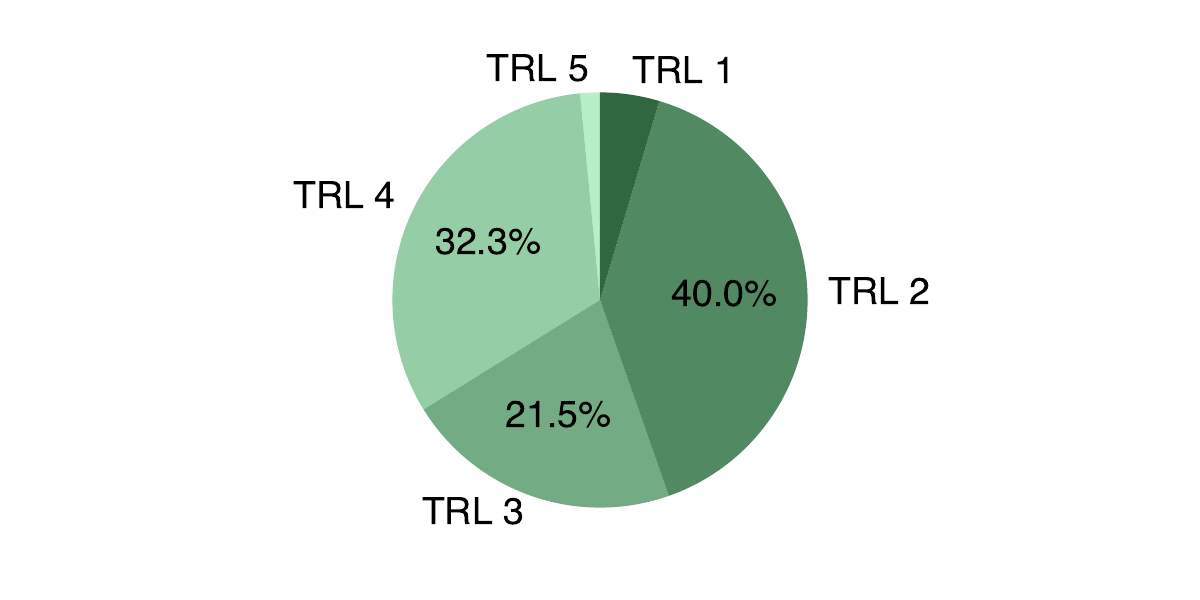}
        (b)
    \end{minipage}
    \caption{Coverage of (a) source code availability and (b) TRL in the selected studies.}
    \label{fig:5w1h:what-others}
\end{figure}

On another front, \Cref{fig:5w1h:what-others}a shows that the vast majority of the existing proposals adopt a \textit{black box} approach.
Namely, most of the existing proposals do not need the source code of software to be available.
This is mainly motivated by the above distribution of proposals, since reference models and good practices (which are the most frequent categories) typically abstract from the source code of a software.
Another motivation comes from the black-box software improvement techniques \cite{03_Karunakaran2013_PetriNet,15_Heithoff2023_DigitalTwins,21_Beghoura2014_GreenEvaluation,26_Jebraeil2017_gUML,30_Alofi2022_OptimizingConsensus,31_Alofi2023_Self-Optimizing,37_Tee2014_GreenKnowledge,40_Wei2023_GreenCodeGeneration,41_Zhang2023_NetAI,57_Lago2019_DecisionMaps,59_Lasla2022_GreenPoWAE,65_Kocak2019_UtilityModel}, which choose a black-box approach to increase their applicability, namely to work also with software whose sources are not available, \eg for privacy or legacy reasons. 
At the same time, the existing white-box proposals show concrete potentials to enhance the environmental sustainability of the software under design/development, as witnessed, \eg by the green code smells detected by the ecoCode project \cite{51_LeGoaer2021_greenCode, 13_LeGoaer2023_ecoCodeProject, 49_LeGoaer2023_ecoCode}.
Following this line, it may be worthy to consider further investigating the potentials of white-box analysis, as well as finding trade-offs between proposals' applicability and enhancements in terms of sustainability.

Finally, \Cref{fig:5w1h:what-others}b displays the overall low TRL of the state of the art on designing and developing environmentally sustainable software.
Most of the studies feature a TRL of 2 or 3, which -- together with TRL 1 -- cover around 66\% of the selected studies.
Among those, only 21.5\% of the selected studies went beyond observing the basic principles and formulating the technology concepts, by proposing an experimental proof-of-concept.
32.3\% of the selected studies went a step higher, by validating their proposal in laboratory and, therefore, reaching TRL 4.
Only the experience report by Mehra et al. \cite{08_Mehra2023_AssessRefactoring} reached TRL 5, being it a report on an industry-relevant case study. 
Overall, the above described distribution of TRLs witnesses a need for increasing the TRL of what is available to support the design and development of environmentally sustainable software, especially if we wish to support the software industry in this task. 

\begin{highlights}
    \highlight{4}{Most of the proposals are reference models, witnessing the need for clarifying what is the environmentally sustainable software design and development}
    \highlight{5}{Existing reference models differ one another, calling for standardisation efforts}
    \highlight{6}{Black-box analyses are the most used, motivated by their applicability. White-box analyses show potentials and should be investigated further, along with trade-offs between applicability and sustainability enhancements}
    \highlight{7}{The overall TRL is low and should be increased to favour the adoption of existing solutions by the software industry}
\end{highlights}

\subsection{Why?}
\label{sec:5W1H:why}
\begin{figure}
    \centering
    \includegraphics[width=.92\textwidth,trim={0 1cm 0 1cm}]{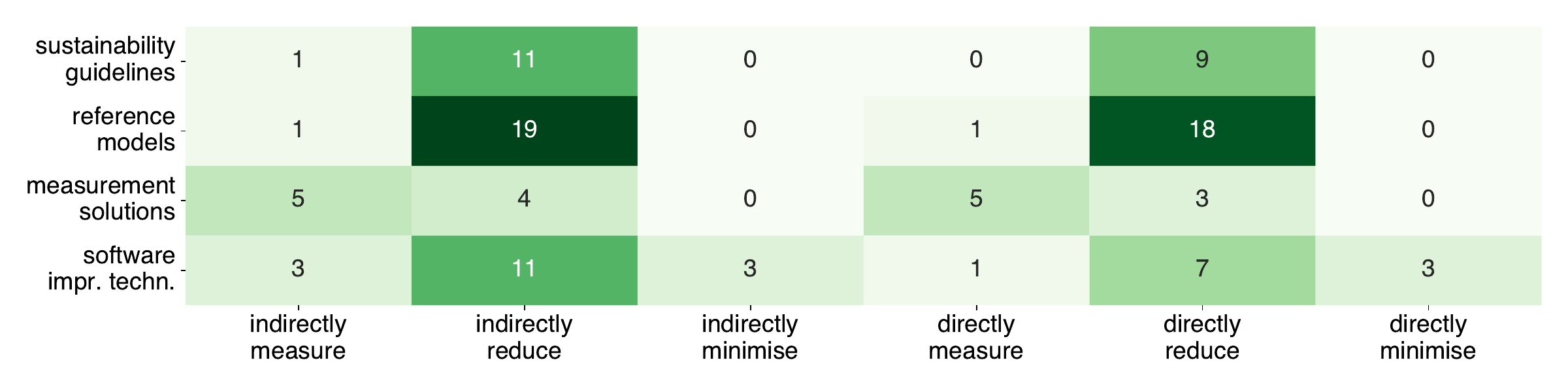}
    \caption{Frequency of energy- and carbon-related goals in the selected studies, distinguished by proposal. Since some studies pursue multiple goals, the sum of frequencies is higher than the number of selected studies.}
    \label{fig:5w1h:heat-why-what}
\end{figure}

The \textit{why} question aims at identifying the sustainability goals in the selected studies, namely whether they support the design and development of \rev{carbon-efficient software by directly or indirectly considering its carbon footprint}.
As shown in \Cref{fig:5w1h:heat-why-what}, the most frequent trends are \rev{reducing the carbon footprint of software through a hybrid approach, namely directly considering a software's carbon footprint, as well as considering the indirect effect of its energy consumption on such footprint}. 
%
%
This is motivated by the types of proposals targeting such goals, \ie reference models and sustainability guidelines provide high level information, which can effectively be exploited to reduce the energy consumption or carbon footprint of software -- but not enough detailed to measure or minimise them.

There are also solutions proposed to enable \rev{directly/indirectly measuring carbon emissions}, which are among the goals of ten and seven of the selected studies, respectively. 
The underlying idea is in line with that of the studies targeting the reduction of carbon emissions.
More precisely, the solutions aimed at \rev{directly/indirectly measuring carbon emissions} are intended to enable comparing the carbon emissions before/after updating a software, with the ultimate goal of assessing whether any applied or planned change contributes to reducing them \cite{53_Patterson2022_MLCO2footprint,54_Kern2013_GreenSA,59_Lasla2022_GreenPoWAE,62_Kipp2011_GreenMetrics}. 

Only three selected studies \cite{31_Alofi2023_Self-Optimizing,30_Alofi2022_OptimizingConsensus,41_Zhang2023_NetAI} pursue the goal of minimising the carbon footprint of software, \rev{doing this with a hybrid approach}.
Such a low number is not only motivated by the type of proposals, as highlighted above, but also by the inherent complexity of optimising by considering the multiple different aspects that influence the carbon footprint of the software under design/development \cite{19_Mancebo2021_FEETINGS,23_Siegmund2022_GreenConfiguration,27_Wedyan2023_Testing,37_Tee2014_GreenKnowledge,38_Sharma2022_GreenQuotient}.
Indeed, Alofi et al. \cite{30_Alofi2022_OptimizingConsensus,31_Alofi2023_Self-Optimizing} and Zhang et al. \cite{41_Zhang2023_NetAI} pursued their optimisation goals by narrowing the search space by relying on the context given by blockchains and ML, respectively.
These are quite recent efforts, which anyhow showed the feasibility of minimising a software's carbon footprint.
Further investigation is needed along this line, possibly also considering, \eg heuristics to achieve sub-optimal minimisation, if the holistic minimisation would get too complex to achieve.

\begin{figure}
    \centering
    \includegraphics[width=.92\textwidth,trim={0 1cm 0 1cm}]{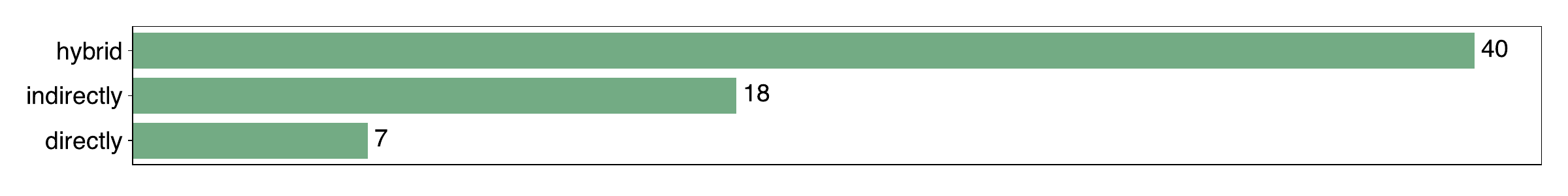}
    \caption{Frequency of pursued goals in the selected studies.}
    \label{fig:5w1h:why-energy-co2}
\end{figure}

Other interesting observations come from \Cref{fig:5w1h:why-energy-co2}, which shows that only seven out of the 65 selected studies consider only the carbon footprint of software.
Indeed, most of the selected studies \rev{pursue their carbon efficiency goals with an indirect or hybrid approach (in 18 and 40 of the selected studies, respectively)}.
In both cases, the carbon footprint is estimated starting from energy consumption while pursuing the goal of measuring/reducing/minimising a software's carbon footprint.
This is mainly motivated by an assumption shared by the selected studies considering energy consumption, namely that reducing the energy consumption of software also reduces its carbon footprint.
However, as demonstrated by Aiello et al. \cite{Aiello2024_CarbonAwareJobScheduling}, energy mixes play a crucial role to relate energy consumption with carbon footprint, as different energy mixes have different associated carbon intensities.
This may negatively impact on the effectiveness of considering energy consumption to indirectly reduce carbon emission.
For example, when a software is powered with clean and renewable energy, there may be no need for reducing its energy consumption, as this may not significantly change its carbon footprint \cite{Forti2024_Carbonstat}.
This calls for further research along the line of directly considering energy mixes and carbon footprint, rather than as an indirect consequence of energy consumption.

\begin{highlights}
    \highlight{8}{The most pursued goals are reducing carbon emissions. Their minimisation can anyhow be achieved by relying on the application domain to narrow the search space of the minimisation problem}
    \highlight{9}{Reducing \rev{energy consumption to indirectly reduce a software's carbon footprint} may not be enough, as different energy mixes may result in different carbon emissions. This calls for solutions directly considering energy mixes and carbon footprint as well}
\end{highlights}

\subsection{Where?}
\label{sec:5W1H:where}
We hereby analyse the application domains, if any, targeted by the existing solutions for the design and development of environmentally sustainable software.
\begin{figure}
    \centering
    \includegraphics[width=.92\textwidth,trim={0 1cm 0 1cm}]{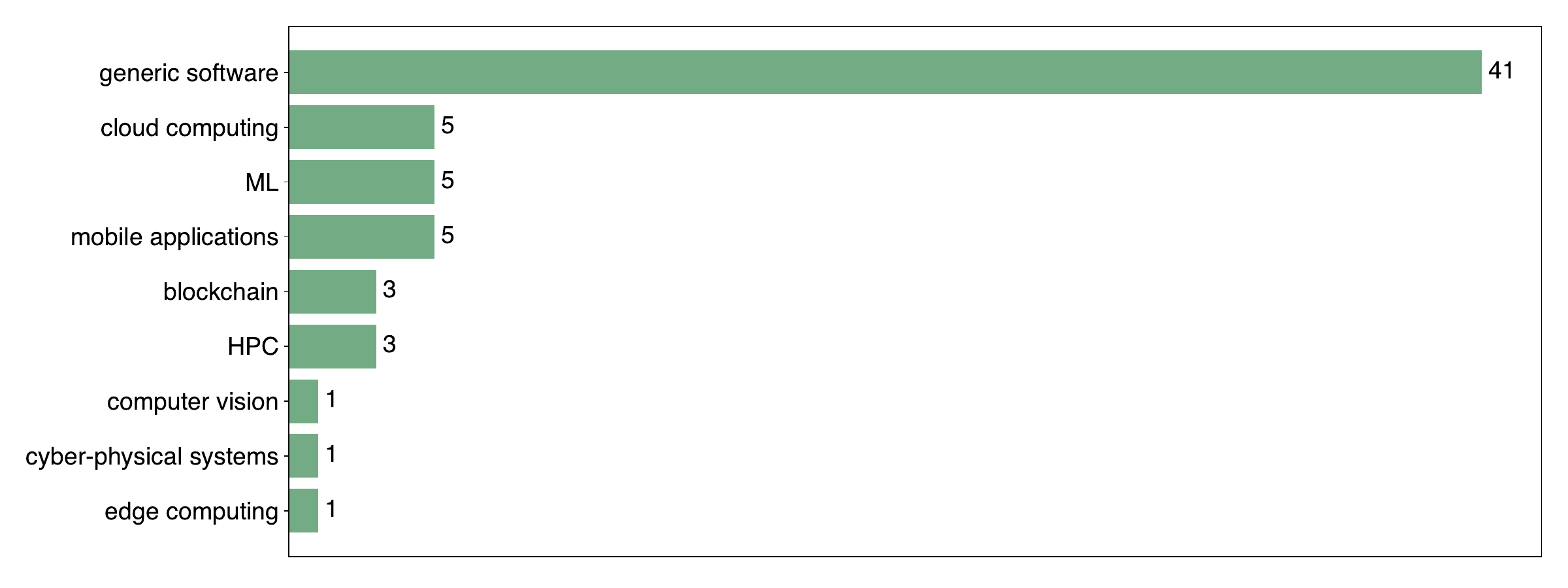}
    \caption{Frequency of application domains in the selected studies.}
    \label{fig:5w1h:where-app-domain}
\end{figure}
\Cref{fig:5w1h:where-app-domain} displays the frequency of application domains in the selected studies, outlining that the vast majority of existing solutions target \textit{generic software} -- rather than application domain-specific software.
\begin{figure}
    \centering
    \footnotesize
    \includegraphics[width=.92\textwidth]{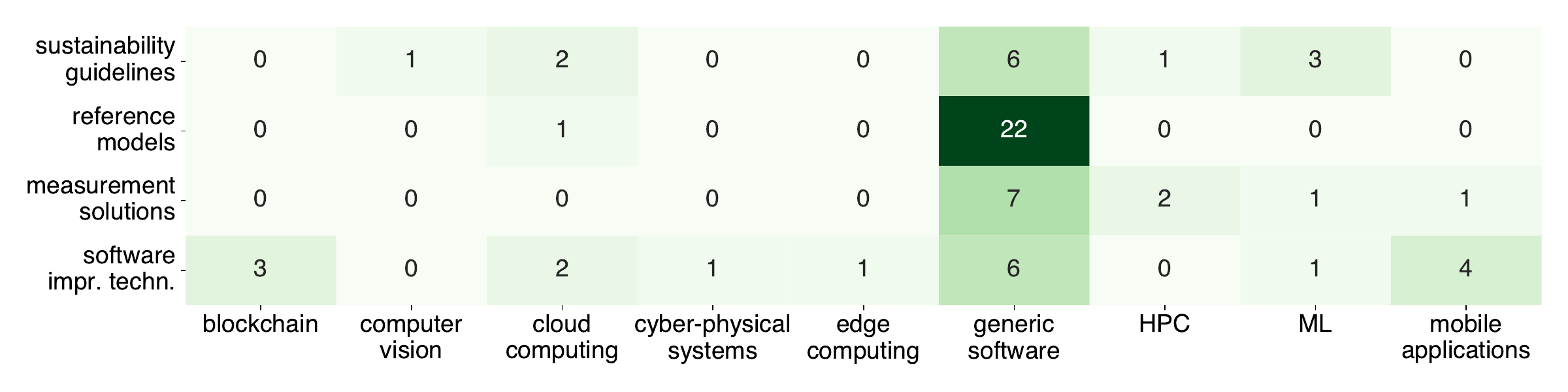}
    \\ (a) \\
    \includegraphics[width=.92\textwidth]{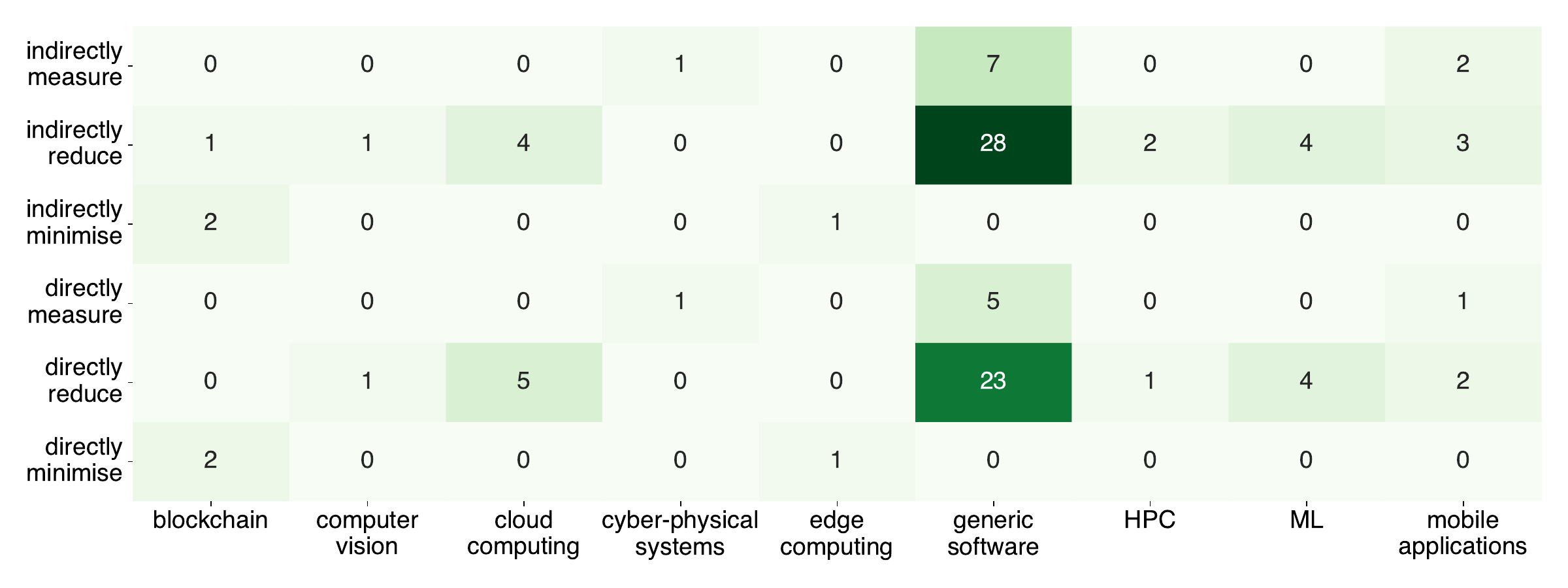}
    \\ (b)
    \caption{Frequency of application domains in the selected studies, distinguished by (a) proposal and (b) goal. As studies can pursue multiple goals, the sum of frequencies in (b) is higher than the number of selected studies.}
    \label{fig:5w1h:heat-where-what}
    \label{fig:5w1h:heat-where-why}
\end{figure}
\rev{This is mainly motivated by the fact that \textit{sustainability guidelines}, \textit{reference models}, and \textit{measurement solutions} are given as high-level, broadly-applicable approaches to reduce the carbon emissions of software, as it can be observed from the heatmaps in \Cref{fig:5w1h:heat-where-what}. The applicability and extension of such generic approaches to specific application domains is, therefore, deserving further investigation, as they might require some adaptation for getting applied to application domain-specific software. The same does not hold for the available \textit{software improvement techniques}, as most of them target specific application domains (\Cref{fig:5w1h:heat-where-what}), rather exploiting the peculiarities of application domain-specific software for reducing its carbon emissions.}

Going beyond generic software, the most targeted application domains are \textit{cloud computing}, \textit{ML}, and \textit{mobile applications}, followed by \textit{blockchain} and \textit{HPC}.
Only single spotted contributions, instead, target the domains of \textit{computer vision}, \textit{cyber-physical systems}, and \textit{edge computing}.
Most of the efforts are indeed geared towards reducing the environmental impact of computationally intensive application domains, which often come with significant energy consumption and carbon footprint \cite{14_Pan2022_CNNDecomposition,41_Zhang2023_NetAI,62_Kipp2011_GreenMetrics}. 
This holds for all application domains mentioned above, but for the case of \textit{mobile applications}, where the focus is more on limiting the energy draining of mobile devices.
The idea is essentially that, even if a small amount of energy is saved on a mobile device, such an energy save have to be summed over billions of devices, and this can contribute to reducing the overall carbon footprint of \textit{mobile applications} \cite{51_LeGoaer2021_greenCode}.

Despite the goal for application domain-specific software is still on reducing their energy consumption and carbon footprint, the way to achieve such goals differs from generic software (\Cref{fig:5w1h:heat-where-what}).
More precisely, only \textit{ML} and \textit{computer vision} are still dealt with high-level sustainability guidelines aimed at reducing their energy consumption and carbon footprint (\Cref{fig:5w1h:heat-where-what}), especially in the model training phase \cite{17_Brownlee2021_AccuracyEnergyML,14_Pan2022_CNNDecomposition,52_Fu2021_ComputerVision,53_Patterson2022_MLCO2footprint,40_Wei2023_GreenCodeGeneration}.
For the other considered application domains, instead, the heat goes over software improvement techniques (\Cref{fig:5w1h:heat-where-what}a), which exploit the application domain-specific context to provide more concrete solutions aimed at reducing their energy consumption and carbon footprint (\Cref{fig:5w1h:heat-where-why}b).
This highlights the potentials of narrowing the scope to specific application domains for providing more concrete techniques to design and develop environmentally sustainable software, hence calling for further investigation along this line.

\begin{highlights}
    \highlight{10}{Most of the existing solutions are high-level guidelines and reference models targeting generic software. Their applicability and extension to specific application domains deserves further investigation}
    \highlight{11}{The currently targeted application domains are those known to be computationally intensive (\eg ML and blockchains) or largely used (\eg mobile applications)}
    \highlight{12}{Narrowing the scope to specific application domains enables providing more concrete techniques to design and develop \rev{carbon-efficient} software, calling for further investigation along this line}
\end{highlights}

\subsection{When?}
\label{sec:5W1H:when}
\begin{figure}
    \centering
    \includegraphics[width=.92\textwidth,trim={0 1cm 0 1cm}]{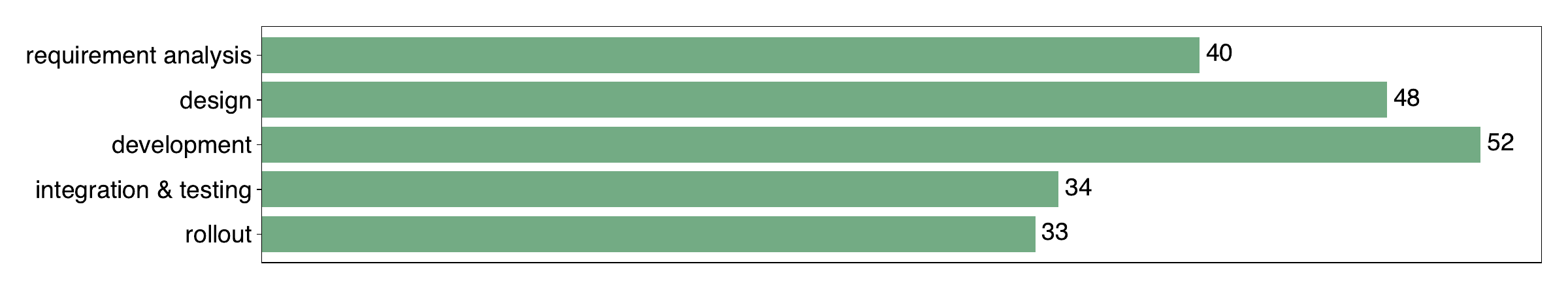}
    \caption{Frequency of lifecycle stages in the selected studies. Since some studies target multiple stages, the sum of frequencies is higher than the number of selected studies.}
    \label{fig:5w1h:when-sdlc}
\end{figure}

\Cref{fig:5w1h:when-sdlc} shows the frequency of lifecycle stages targeted in the selected studies. 
Being the present SLR targeting the \textit{design} and \textit{development} of environmentally sustainable software, such stages naturally result to be the most targeted by the selected studies.
The other three stages are anyhow quite frequent in the selected studies, with \textit{requirement analysis} targeted by 40 studies, \textit{integration \& testing} targeted by 34 studies, and \textit{rollout} targeted by 33 studies.
\rev{This is mainly due to the fact that the currently available \textit{reference models} and \textit{measurement solutions} -- which together account for more than half of the selected studies -- are designed to be applicable across all lifecycle stages, thereby naturally extending their relevance beyond design and development.}
\begin{figure}
    \centering
    \footnotesize
    \includegraphics[width=.92\textwidth]{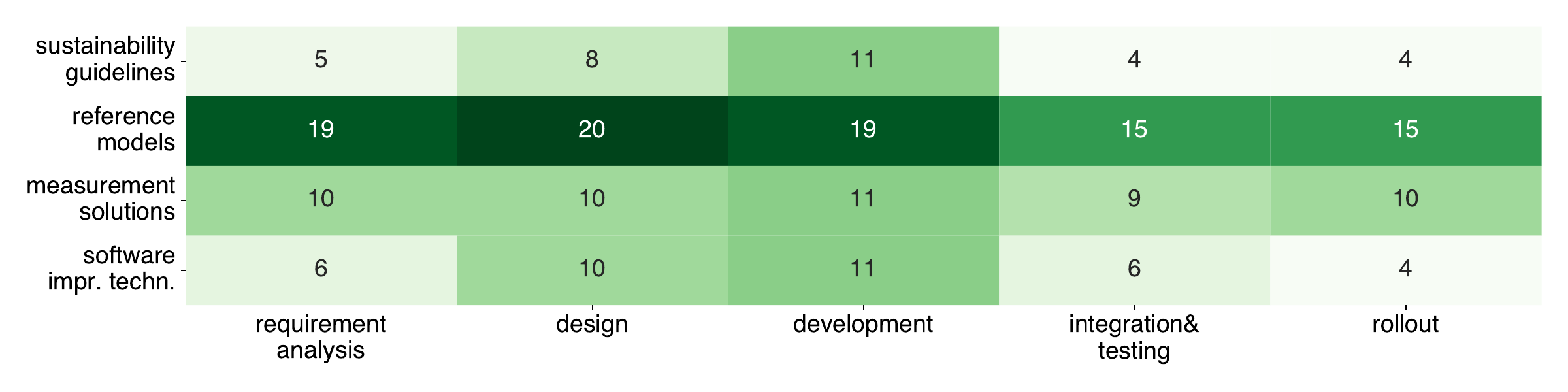}
    \\ (a) \\ 
    \includegraphics[width=.92\textwidth]{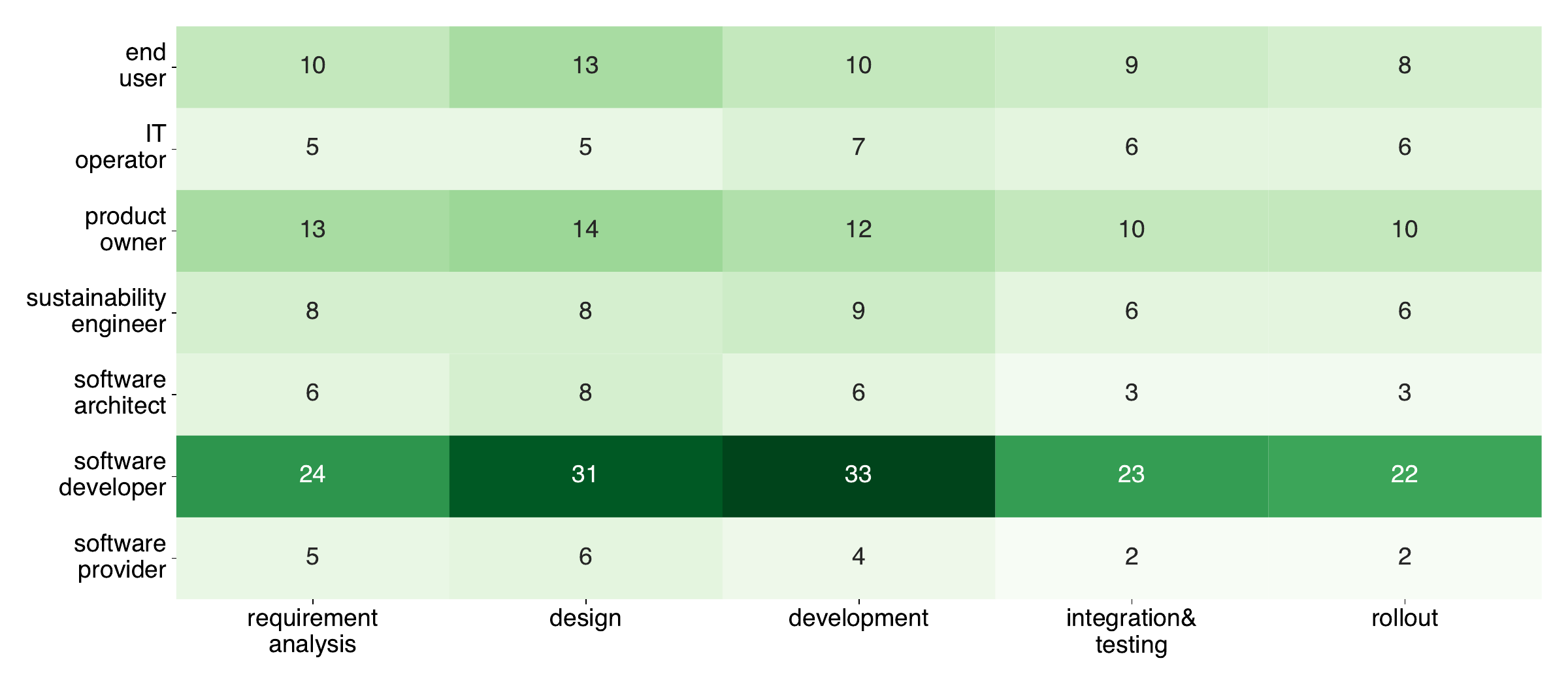}
    \\ (b) \\ 
    \caption{Frequency of lifecycle stages in the selected studies, distinguished by (a) proposal and (b) stakeholder. Since some of the selected studies target multiple stakeholders/stages, the sum of frequencies is higher than the number of selected studies.}
    \label{fig:5w1h:heat-when-others}
\end{figure}
%
\rev{In contrast, sustainability guidelines and software improvement techniques tend to be more concentrated on these two stages, with less than half addressing other phases of the lifecycle}
(\Cref{fig:5w1h:heat-when-others}).

On another front, it is interesting to observe who is involved in which lifecycle stage, according to the state of the art on the design and development of environmentally sustainable software. 
\Cref{fig:5w1h:heat-when-others}b provides such information, confirming what we observed in \Cref{sec:5W1H:who}: the most involved stakeholders are \textit{software developers}. 
Notably, \textit{software developers} are not only involved in the \textit{development} stage, but rather result the most involved across all the stages of a software's lifecycle -- which is again in line with what we observed in in \Cref{sec:5W1H:who}, namely that the burden of environmental sustainability is expected to be put on \textit{software developers}.
This, along with the fact that the spanning across all stages seems to apply to all stakeholders, suggests that all stakeholders should be involved -- and supported -- during all lifecycle stages.
Therefore, further investigation should be pursued along this line.

\begin{highlights}
    \highlight{13}{All lifecycle stages result to be significantly covered, despite our focus is on \textit{design} and \textit{development}}
    \highlight{14}{Reference models and measurement solutions span across all lifecycle stages, whereas sustainability guidelines and software improvement solutions focus more on design and development only}
    \highlight{15}{Software developers are the most involved in all stages, but all stakeholders span across all lifecycle stages, calling for supporting all stakeholders during all lifecycle stages}
\end{highlights}

\subsection{How?}
\label{sec:5W1H:how}
\begin{figure}
    \centering
    \includegraphics[width=.92\textwidth,trim={0 1cm 0 1cm}]{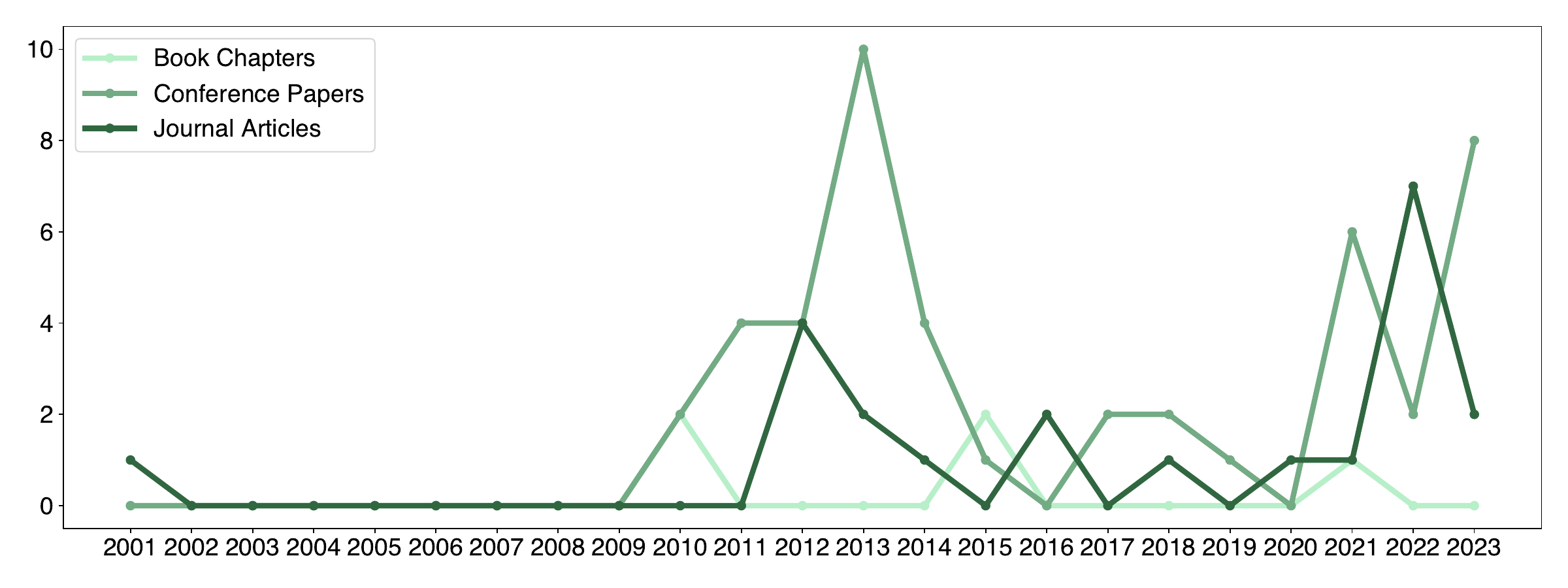}
    \caption{Yearly numbers of publications (\ie \textit{book chapter}s, \textit{conference paper}s, and \textit{journal article}s) related to designing and developing environmentally sustainable software.}
    \label{fig:5w1h:how-year}
\end{figure}

The \textit{how} question aims at identifying the distribution of publications on environmentally sustainable software design and development, based on bibliometrics like the publication type and year (\Cref{fig:5w1h:how-year}).
After a pioneering journal article in 2001, no activity is reported up to 2009 included.
Instead, since 2010 onwards, there were publications on the topic each year, witnessing a renewed interest in the topic, with a peak in 2013 and an ongoing growth since 2020.
Overall, conference papers were the most used means of publication. 
This may be motivated by the authors wishing to timely publish their results, therefore going for conferences that offer a faster publication. 
An anomaly in this respect is 2022, with more journal articles than conference papers, which may be motivated by the faster publication process recently adopted by international journals.

Notably, the peak in 2013 and the ongoing growth seem to relate two co-occurring events in time.
More precisely, the peak in 2013 may be a consequence of a report \cite{OECD2009_Report} by OECD (\textit{Organisation for Economic Cooperation and Development}), which assessed government programmes and business initiatives on ICT and environmental sustainability across 22 OECD countries.
Instead, the ongoing growth since 2020 may be a consequence of the recent interest in environmental sustainability.
These were raised, \eg by international organisations like Fridays for Future, and investment policies like the European strategic agenda \cite{EU2019_StrategicAgenda}, which calls for and finances a sustainable growth in European industries, including the IT industry \cite{EU2019_GreenDeal,EU2021_FinancingClimateTransition,EU2024_IndustrialPolicy}. 


\begin{highlights}
    \highlight{16}{Conference papers are steadily higher in number than journal articles, mainly because of conferences' faster publication tracks}
    \highlight{17}{The peak of publication in 2013 may be motivated by the 2009 OECD assessment of government programmes and business initiatives on ICT and environmental sustainability}
    \highlight{18}{An ever \rev{increasing} number of publications since 2020 witnesses the hotness of the topic, also pairing with the ever increasing interest in environmental sustainability}
\end{highlights}

\section{Threats to Validity}
\label{sec:ttv}
Based on the taxonomy developed by Wohlin et al. \cite{Wohlin2000_ExperimentationSE}, four potential threats may apply to the validity of our study.
These are the threats to \textit{external}, \textit{internal}, \textit{construct}, and \textit{conclusions} validities, which we discuss hereafter.

\smallskip \noindent 
\textbf{Threats to External Validity}.
The potential threats to the external validity of a study affect the applicability of the study's results in a broader and more general context \cite{Wohlin2000_ExperimentationSE}.
In our case, since the selected studies are obtained from a large extent of online available sources, our results and observations may be only partly applicable to the broad area of the design and development of \rev{carbon efficient} software, and this may potentially threaten the external validity of our study.
To mitigate this potential threat, we organised multiple feedback sessions during our study. We analysed the discussion following up from each feedback session, and we exploited the resulting qualitative data to fine-tune both our research methods and the applicability of our findings. 
We also prepared a replication package bundling the intermediary artefacts and the final results produced while running our study, which is publicly available online \cite{Danushi2024_ReplicationPackage}.
The replication package, together with the references to the selected studies listed in \Cref{tab:design:selected-studies}, can be used to deepen the understanding of our data, as well as to repeat our study and verify our findings.

Additionally, the external validity of our study may be threatened by the risk of having missed relevant studies, since the concepts related to those included in our search strings are differently named in such studies.
To mitigate this potential threat, we have explicitly included all relevant synonyms in our search strings and we also enacted backward snowballing (see \Cref{sec:design:search}).

\smallskip \noindent 
\textbf{Threats to Internal Validity}
The internal validity of a study concerns the validity of the method employed to analyse the study data \cite{Wohlin2000_ExperimentationSE}.
To reinforce the internal validity of our study, we relied on established guidelines to conduct systematic literature reviews in software engineering \cite{Kitchenham2007_GuidelinesSLRinSE,Petersen2008_SMSGuidelines,Petersen2015_SMSGuidelinesUpdate}.
Additionally, we enacted various triangulation rounds and adopted inter-rater reliability assessment (see \Cref{sec:design}), which were conceived to avoid possible biases by triangulation.

\smallskip \noindent 
\textbf{Threats to Construct and Conclusions Validity}
The construct and conclusions validities concerns the generalisability of the constructs under study and the degree to which the conclusions of a study are reasonably based on the available data, respectively \cite{Wohlin2000_ExperimentationSE}.
The construct and conclusions validities of our study may have been affected by observer and interpretation biases.
To mitigate these potential threats, we exploited inter-rater reliability assessment (see \Cref{sec:design}), and we also independently drawn conclusions from the available data.
Such conclusions were then suitably merged and double-checked against the selected (and related) studies in a joint discussion session.

\section{Related Work}
\label{sec:related}
Several secondary studies survey the state of the art for sustainable software.
A first example is given by Ormazabal et al \cite{Ormazabal2014_CarbonFootprintSoftware}, who survey the software tools that can be used to assess a product's lifecycle and analyse its carbon footprint.
Sinha and Chandel \cite{Sinha2014_SoftwareToolsHybridRenewableEnergySystems} and Sudarmadi and Garniwa \cite{Sudarmadi2023_TechnoEconomicAnalysisHomer} review the use of software tools for planning of hybrid renewable energy systems.
Popescu et al.~ \cite{Popescu2022_DigitalTwinsForEnvironmentalProduction} survey the existing approaches for building digital twins of manufacturing production systems by incorporating environmental sustainability aspects.
The above listed studies differ from ours in their objective, as they discuss how to achieve environmental sustainability \textit{with} software.
We instead focus on the solutions improving the environmental sustainability \textit{of} software itself.

Other noteworthy secondary studies are the literature reviews by Salam and Khan \cite{Salam2016_GreenSuccessFactors} and Sambhanthan and Potdar \cite{Sambhanthan2016_BusinessSustainabilitySurvey}.
Salam and Khan \cite{Salam2016_GreenSuccessFactors} classify and analyse the success factors for vendors of green software.
Sambhanthan and Potdar \cite{Sambhanthan2016_BusinessSustainabilitySurvey} instead focus on the existing frameworks for assessing a business' sustainability, therein included ICT sustainability. 
Therefore, the reviews by Salam and Khan \cite{Salam2016_GreenSuccessFactors} and Sambhanthan and Potdar \cite{Sambhanthan2016_BusinessSustainabilitySurvey} differ from ours in the objective, as they focus on the business-related aspects of software's sustainability. 
We instead aim at classifying the technical solutions for designing and developing \rev{carbon-efficient} software.

Andrikopoulos et al.~\cite{Andrikopoulos2022_SMSSustainabilitySWArch} provide a systematic mapping study aimed at analysing which architecting activities are affected/affect which of the four different dimensions of sustainability (\ie technical, economic, environmental, and social).
Ahmadisakha and Andrikopoulos \cite{Ahmadisakha2024_ArchitectingSustainabilityCloud} deepen into the topic, focussing specifically on Cloud-based software architecture, still analysing how architectural decisions for Cloud software are affected/affect the four different dimensions of sustainability. 
Therefore, the scope and goal of both the above secondary studies differ from ours, as they aim at eliciting which of the four different dimensions of sustainability impact on/are impacted by which architecting phase.
We instead focus on the 5W1H of designing and developing \rev{carbon-efficient} software.
Additionally, Ahmadisakha and Andrikopoulos \cite{Ahmadisakha2024_ArchitectingSustainabilityCloud} focus on Cloud-based software, while we rather consider multiple different application domains.

Other secondary studies focus on improving the environmental sustainability of software in specific target application domains.
For example, Azimzadeh and Tabrizi \cite{Azimzadeh2015_TaxonomySurveyGreenDataCenters} and Khan et al.~\cite{Khan2020_SoftwareTechniquesGreenDataCenters} provide two different reviews on software techniques for managing green Cloud data centres.
Gaglianese et al.~\cite{Gaglianese2023_GreenCloudEdgeOrchestration} survey the existing approaches for the environmentally sustainable orchestration of containerised software applications over Cloud-Edge infrastructures.
Casta\~{n}o et al.~\cite{Castano2023_CarbonFootprintHuggingFaceMLModels} analyse and classify the carbon footprint of training publicly available ML models.
Verdecchia et al.~\cite{Verdecchia2023_GreenAIReview} review the state of the art on green AI, by distinguishing the proposals supporting the training and operation of AI models.
L{\'{o}}pez{-}P{\'{e}}rez et al.~\cite{LopezPerez2022_5GRadioAccessNetworkEnergyEfficiencySurvey} survey the existing techniques for operating 5G networks in an energy-efficient manner.
Thus, all the above-mentioned secondary studies focus on a specific application domain and on different lifecycle stages than ours, as we rather consider designing and developing \rev{carbon-efficient} software by covering -- and classifying -- multiple different application domains.

The SLRs by Bozzelli et al.~\cite{Bozzelli2014_GreenSoftwareMetricsReview}, Calero et al.~\cite{Calero2013_SLRSoftwareSustainabilityMeasures}, Koziolek\cite{Koziolek2011_SustainabilitySWArchitectures}, and Venters et al.~\cite{Venters2018_SLRSustainableSE,Venters2023_SLRSustainableSEFollowUp} are a step closer to ours, as they all pertain to the broader scopus of environmental sustainability in software engineering.
More precisely, Bozzelli et al.~\cite{Bozzelli2014_GreenSoftwareMetricsReview} and Koziolek \cite{Koziolek2011_SustainabilitySWArchitectures} classify the existing metrics for estimating the greenness of software.
Calero et al.~\cite{Calero2013_SLRSoftwareSustainabilityMeasures} elicit the software quality aspects related to environmental sustainability.
Venters et al.~\cite{Venters2018_SLRSustainableSE,Venters2023_SLRSustainableSEFollowUp} go a step further, by also identifying which definitions of software sustainability are available, as well as how software sustainability impacts on architectural decisions, goals, and education.
Thus, even if slightly overlapping, the objectives of the above mentioned SLRs differ from ours, as we rather aim at answering to the 5W1H questions for designing and developing \rev{carbon-efficient} software (see \Cref{sec:design:rq}).

Similar arguments apply to the SLR by Penzenstadler et al.~\cite{Penzenstadler2012_SLRSustainableSE}, who started with similar aims than ours, to then realise that -- at the time when the study was conducted -- only a few studies tackled environmental sustainability in software engineering.
The SLR by Penzenstadler was then redirected to identify and classify the broader scope of relating environmental sustainability to software, one of these being how it relates with software engineering.
Therefore, the SLR by Penzenstadler et al.~\cite{Penzenstadler2012_SLRSustainableSE} differs from ours in the ultimate contribution, as -- after more than 10 years -- the literature on sustainability is more mature, as we showed by focusing on the narrower scope of designing and developing \rev{carbon-efficient} software.

In summary, to the best of our knowledge, ours is the first SLR on the state of the art on designing and developing \rev{carbon-efficient} software.
It is the first doing it considering not only the \textit{what} of sustainable software design and development, but rather aiming at answering to the all 5W1H questions related to designing and developing \rev{carbon-efficient} software (see \Cref{sec:design:rq}).
This enabled us, \eg to first reason on the stakeholders for 
\rev{carbon-efficient} software design and development, to span over multiple application domains, and to first distinguish whether environmental sustainability is dealt with by only reducing energy consumption or by also considering the carbon footprint of software.

\section{Conclusions and Outlook}
\label{sec:conclusions}
We presented the results of a SLR on the design and development of 
\rev{carbon-efficient} software.
After classifying and recapping the existing solutions for the tasks, we analysed \textit{who} are the targeted stakeholders, \textit{what} methods allow to design and develop 
\rev{carbon-efficient} software, \textit{why}, \textit{where} and \textit{when} the proposed solutions are applied (in terms of application domains and lifecycle stages, respectively), and \textit{how} the selected studies have been presented to the scientific community.

The results of our SLR can help researchers and practitioners identify and adopt existing guidelines and reference models for designing and developing 
\rev{carbon-efficient} software, as well as to select applicable solutions for their software projects. Our analysis also highlights a growing interest in this topic since 2020, providing a basis for establishing future research directions. To facilitate this, we outline seven research directions derived from our analysis, each mapped to the highlights in \Cref{sec:5w1h} that motivate these directions.
\begin{itemize}
\item
\textit{Relieving the burden from software developers} ($H_1$--$H_3,H_{15}$). 
The existing solutions mainly aim at supporting software developers, and this somehow puts the burden of designing and developing 
\rev{carbon-efficient} software on software developers themselves.
There is a need for solutions \rev{to} support and involve all the potential stakeholders in the process.
In particular, we should enable software architects and sustainability engineers to get more involved in the design of 
\rev{carbon-efficient} software, taking responsibility for sustainability-oriented design choices, thus complementing the work and efforts carried out by software developers.

\item
\textit{Standardising \rev{carbon efficiency} for software} ($H_4,H_5$).
Most of the surveyed proposals are reference models, witnessing the need for clarifying the notion of \rev{carbon efficiency} for software, also outlining how this can be used to support the design and development of more sustainable software.
However, despite sharing some base concepts, the currently available reference models differ one from another, calling for standardisation efforts.
The availability of a standardised reference model for software \rev{carbon efficiency} would enable all stakeholders to rely on the same understanding of \rev{carbon efficiency}. 
It would also lay the foundations for devising full-fledged solutions to support the design and development of 
\rev{carbon-efficient} software.

\item
\textit{Finding a trade-off between applicability and \rev{carbon efficiency}} ($H_6$). 
Most of the existing solutions for designing and developing 
\rev{carbon-efficient} software adopt a black-box approach.
This is mainly motivated by applicability reasons, as black-box approaches can work even with software whose sources are not available, \eg for privacy or legacy reasons.
At the same time, while surveying the state of the art, we observed some potentials of white-box analyses.
For example, Le Goa\"{e}r et al \cite{51_LeGoaer2021_greenCode,13_LeGoaer2023_ecoCodeProject,49_LeGoaer2023_ecoCode} highlighted the potential of defining, detecting, and resolving so-called green code smells, namely possible symptoms of bad coding decisions that may worsen the environmental impact of software. 
Mehra et al. \cite{08_Mehra2023_AssessRefactoring} instead showcased how static analysis enabled them to successfully resolve code inefficiencies in an industrial use case. 
Therefore, the potentials of white-box approaches deserve further investigation, in an effort to find a suitable trade-off between the applicability of a solution and its effectiveness in supporting the design and development of \rev{carbon-efficient} software.

\item
\textit{Raising the overall TRL} ($H_7$). 
We observed that, overall, the TRL of the state of the art on 
\rev{carbon-efficient} software design and development is relatively low (with only one out of 65 selected studies reaching a TRL 5).
To favour the adoption of existing -- and newly proposed -- solutions by the software industry, the overall TRL should be increased.
In particular, the existing proposals should be prototyped and assessed over industry-relevant scenarios, to then release more mature tooling to the software industry.
To this end, there is also a need for benchmarks, industry-relevant use cases, and realistic case studies that could be used to thoroughly validate existing/newly proposed solutions for designing and developing 
\rev{carbon-efficient} software.

\item
\textit{Increasing carbon-awareness} ($H_9$).
Most of the existing solutions consider energy consumption as the main discriminant to improve the environmental sustainability of software.
This is motivated by the shared assumption that, by directly reducing the energy consumption of software, its carbon footprint gets indirectly reduced as well.
However, energy mixes play a crucial role to relate energy consumption with carbon footprint, as different energy mixes come with different associated carbon intensities \cite{Aiello2024_CarbonAwareJobScheduling}.
This may impair the effectiveness of directly considering energy consumption to indirectly reduce carbon emission, since \eg 
when software is powered with clean and renewable energy, reducing its energy consumption does not significantly change its carbon footprint \cite{Forti2024_Carbonstat}.
This calls for further research along the line of directly considering energy mixes and carbon emissions of software, rather than only focussing on reducing its carbon footprint by limiting its energy consumption.

\item
\textit{Adopting existing solutions in application domains} ($H_8$,$H_{10}$--$H_{12}$). 
Most of the existing solutions are high-level guidelines, reference models, and measurement solutions for generic software.
Being them generic, their concrete adoption in specific application domains may require considerable efforts, \eg to adapt them to work with the peculiar aspects that mostly impact on the carbon footprint on application domain-specific software (like the training phase for ML models, or the PUE for Cloud data centers).
This calls for identifying and considering such peculiar aspects, to define/improve domain-specific guidelines and reference models, as well as to adapt existing techniques to work with application domain-specific software.
This would enable the effective use and assessment of existing solutions available in multiple different application domains, thus possibly paving the way towards their adoption.
Additionally, narrowing the scope to specific application domains showed the potential to not only reduce, but also minimise, the \rev{carbon efficiency} of the software under design/development.

\item
\textit{Relating design and development with all lifecycle stages} ($H_{13}$--$H_{15}$). 
Even if focussing on the design and development of 
\rev{carbon-efficient} software, this SLR highlighted how the existing solutions span across all the stages of the software lifecycle.
This is motivated by the fact that sustainability guidelines and reference models are often proposed for the full software lifecycle.
Going beyond this, we observed the effectiveness of the design and development stages in achieving 
\rev{carbon-efficient} impacts on/is impacted by the other lifecycle stages as well.
This calls for supporting the software design and development stages by considering, keeping track of, and assessing their impacts/relations with the whole software lifecycle. 
\end{itemize}

\begin{revenv}
Additionally, although this survey focuses specifically on the design and development stages, the outcomes discussed in \Cref{sec:5W1H:when} highlight that other lifecycle stages might contribute to the carbon-efficiency of software artifacts, \eg requirements engineering might set carbon-efficiency goals, testing might help validating carbon-efficiency, and operation might further reduce the carbon footprint of running software.
While the carbon-efficient operation of software has already been analysed, \eg in Gaglianese \etal  \cite{Gaglianese2023_GreenCloudEdgeOrchestration}, future secondary studies could complement the results of this work by exploring how the other lifecycle stages can further contribute to the carbon-efficiency of software artifacts. 
Such secondary studies are left in the scope of future work.
\end{revenv}

\begin{acks}
This work has been partly supported by projects: \textit{FREEDA} (CUP: I53D23003550006), funded by the frameworks PRIN (MUR, Italy) and Next Generation EU and \textit{Energy-aware management of software applications in Cloud-IoT ecosystems} (RIC2021\_PON\_A18) funded over ESF REACT-EU resources by the \textit{Italian Ministry of University and Research} through \textit{PON Ricerca e Innovazione 2014--20}. 
\end{acks}
  
\bibliographystyle{ACM-Reference-Format}
\bibliography{src/_biblio,src/_selected-studies}

\end{document}